\documentclass{emulateapj}
\usepackage{graphicx}
\newcommand\msun{\ensuremath{M_\sun}}
\newcommand\teff{\ensuremath{T_{\rm eff}}}
\newcommand\logg{\ensuremath{\log g}}
\newcommand\tcool{\ensuremath{\tau_{\rm cool}}}

\newcommand\minit{\ensuremath{M_{\rm init}}}
\newcommand\mwd{\ensuremath{M_{\rm WD}}}
\newcommand\mcrit{\ensuremath{M_{\rm up}}}
\newcommand\ubv{\ensuremath{U\!BV}}
\newcommand{\ebv}{\ensuremath{E(\bv)}}
\newcommand{\uv}{\ensuremath{U\! - \! V}}

\slugcomment{Accepted for publication in the Astrophysical Journal} 
\shorttitle{White Dwarfs in M35}
\shortauthors{Williams, Bolte, \& Koester}

\begin{document}
\title{Probing The Lower Mass Limit for Supernova Progenitors and the
High-Mass End of the Initial-Final Mass Relation from White Dwarfs in
the Open Cluster M35 (NGC 2168)\altaffilmark{1}}

\author{Kurtis A. Williams\altaffilmark{2}}
\email{kurtis@astro.as.utexas.edu}
\affil{Department of Astronomy, University of Texas, Austin, TX, USA}
\and
\author{Michael Bolte}
\email{bolte@ucolick.org}
\affil{UCO/Lick Observatory, University of California, Santa Cruz, CA,
  USA}
\and
\author{Detlev Koester}
\email{koester@astrophysik.uni-kiel.de}
\affil{Institut f\"ur Theoretische Physik und Astrophysik, University
  of Kiel, Kiel, Germany}

\altaffiltext{1}{Some of the data presented
herein were obtained at the W.M. Keck Observatory, which is operated
as a scientific partnership among the California Institute of
Technology, the University of California and the National Aeronautics
and Space Administration. The Observatory was made possible by the
generous financial support of the W.M. Keck Foundation.}

\altaffiltext{2}{NSF Astronomy \& Astrophysics Postdoctoral Fellow}

\begin{abstract}
We present a photometric and spectroscopic study of the white dwarf
population of the populous, intermediate-age open cluster M35 (NGC
2168); this study expands upon our previous study of the white dwarfs
in this cluster. We spectroscopically confirm 14 white dwarfs in the
field of the cluster: 12 DAs, 1 hot DQ, and 1 DB star.  For each DA,
we determine the white dwarf mass and cooling age, from which we
derive the each star's progenitor mass.  These data are then added to
the empirical initial-final mass relation (IFMR), where the M35 WDs
contribute significantly to the high-mass end of the relation. The
resulting points are consistent with previously-published linear fits
to the IFMR, modulo moderate systematics introduced by the uncertainty
in the star cluster age.  Based on this cluster alone, the
observational lower limit on the maximum mass of white dwarf
progenitors is found to be $\sim 5.1\msun-5.2\msun$ at the 95\%
confidence level; including data from other young open clusters raises
this limit as high as $7.1\msun$, depending on the cluster membership
of three massive WDs and the core-composition of the most massive WDs.
We find that the apparent distance modulus and extinction derived
solely from the cluster white dwarfs ($(m-M)_V=10.45\pm 0.08$ and
$\ebv=0.185\pm 0.010$, respectively) is fully consistent with that
derived from main-sequence fitting techniques.  Four M35 WDs may be
massive enough to have oxygen-neon cores; the assumed core composition
does not significantly affect the empirical IFMR.  Finally, the two
non-DA WDs in M35 are photometrically consistent with cluster
membership; further analysis is required to determine their
memberships.
\end{abstract}

\section{Introduction}

Intermediate-mass stars ($M \sim 6\msun - 10\msun$) end their
evolution in one of two ways, either exploding as a core-collapse
supernova or losing large amounts of material to form a massive white
dwarf (WD) star. Recent models of massive asymptotic giant branch
(AGB) stars suggest that, for stars with initial (i.e., zero-age main
sequence) masses $\lesssim 7.25\msun$, the endpoint of stellar
evolution will be a carbon-oxygen (C/O) WD.  For stars with initial
masses $\sim 7.25\msun - 9.0\msun$, carbon will burn in a
``super-AGB'' star, forming a degenerate oxygen-neon (ONe) core which could
become an ONe WD
\citep[e.g.,][]{1997ApJ...485..765G,2006MmSAI..77..846P,2008ApJ...675..614P}.
More massive super-AGB stars may explode as electron capture
supernovae, and stars with masses $\gtrsim 11\msun$ explode as
canonical core-collapse supernovae.

The initial mass dividing supernova progenitors from white dwarf
progenitors, \mcrit, (also known as $M_w$ and referred to as $M_{\rm crit}$
in our earlier work) is therefore likely to lie in this mass range
($\sim 7\msun - 9\msun$); indeed, current super-AGB models predict
$\mcrit\approx 8-9\msun$, depending on metallicity and the degree of
overshooting \citep{2007A&A...476..893S}.  However, observational
constraints on \mcrit\ have been slow in coming. The best published
observations give $\mcrit=8^{+3}_{-2}\msun$
\citep{1996A&A...313..810K}.

Tight constraints on \mcrit\ are important; for a burst of star
formation with a Salpeter initial mass function
\citep[IMF;][]{1955ApJ...121..161S}, the number of stars in the mass
range of $6\msun - 10\msun$ is equal to the number of stars with
$M\geq 10\msun$.  This factor of two uncertainty in the number of
supernovae in a starburst region has important implications for
quantifying supernova feedback on the interstellar medium
\citep[e.g.,][]{1977ApJ...218..148M}, and for understanding
supernova-driven winds in galaxies
\citep[e.g.,][]{1986ApJ...303...39D,2005ApJ...621..227M}.

Observations of WDs provide a lower limit on the value of \mcrit. In
the simplest form, this limit can be determined by identifying WDs in
progressively younger star clusters until no WDs are found, indicating
that the turnoff-mass stars are going supernova instead of forming WD
remnants.  Variations on this method were used by
\citet{1980ApJ...235..992R} and \cite{1982ApJ...255..245A}, who
independently determined $\mcrit\geq 4\msun-5\msun$.  This method is
improved upon by using spectroscopic analysis of the WDs in each
cluster to determine each WD's progenitor mass.  The most massive WD
progenitor is then a lower limit on \mcrit.  Such analysis was
performed by \citet{1996A&A...313..810K} in the open cluster
NGC 2516, where they found $\mcrit=8^{+3}_{-2}\msun$.  In
our initial study of WDs in M35, we determined
$\mcrit\geq 5.8\msun$ \citep{2004ApJ...615L..49W}.

The relationship of a WD's mass to that of its progenitor, the
initial-final mass relation (IFMR), is also a matter of keen interest.
This relationship quantifies the integrated mass lost by a star over
its entire evolution, and is therefore a necessary part of
understanding chemical enrichment and star formation efficiency in
galaxies \citep[e.g.,][]{2005MNRAS.361.1131F}.  The IFMR represents
one of the best observational constraints on AGB star mass loss.
These stars are thought to be the primary sites of $s$-process
production and to play crucial roles in the abundance ratios of
helium, carbon, and nitrogen \citep[e.g.,][and references
therein]{1999ARA&A..37..239B,2001A&A...370..194M,2005ARA&A..43..435H}.

Due in large part to the steepness of the IMF, the IFMR has relatively
few points in the high initial mass ($\minit\gtrsim 4\msun$) region.
Four massive WDs are known from the open cluster NGC 2516
\citep{1996A&A...313..810K}.  The Pleiades provides one to three
points, depending on whether the massive WDs GD 50 and PG 0136+251
were once cluster members \citep{2006MNRAS.373L..45D}. Age-dating of
Sirius A has allowed the progenitor mass of Sirius B to be
calculated \citep{2005ApJ...630L..69L}. A new paper by
Dobbie2008 identifies three massive WD progenitors in
NGC 3532 and two in NGC 2287.  Finally, the older
open cluster NGC 2099 has one high-mass WD, albeit with very
large error bars on both the initial and final mass
\citep{2005ApJ...618L.123K}.

It is suspected that the IFMR should have some metallicity dependence
\citep[e.g.,][]{2007A&A...469..239M}, with more metal-rich systems
producing lower mass WDs.  Indeed, the WDs in the super-metal-rich
cluster NGC 6791 have surprisingly low masses
\citep{2007ApJ...671..748K}, though the invoked enhanced mass loss due
to the high metallicity results in stars circumventing the AGB phase
altogether.  Comparison of the WD masses in the open star clusters NGC
2099, the Hyades, and Praesepe may show some evidence of metallicity
dependence \citep{2005ApJ...618L.123K}, though this claim hinges on a
yet-unpublished significantly sub-solar metallicity measurement for
NGC 2099.  Published spectroscopic metallicity measurements for this
cluster \citep{2005AJ....130.1916M,2008ApJ...675.1233H} claim a solar
metallicity or slightly super-solar metallicity for NGC 2099; the
younger cluster age resulting from a higher metallicity would erase
most of the claimed signal.  In short, the metallicity dependence of
the IFMR, while expected, has yet to be observed conclusively.

Addressing the questions of \mcrit, the shape of the high-mass end of
the IFMR, and any metallicity dependence via observations thus
requires WD observations in multiple star clusters that are young
(ages $\sim 50-250$ Myr), relatively rich, and of markedly different
metallicities.

\subsection{The Open Cluster M35\label{sec.intro.m35}}
The open cluster NGC 2168 (M35) is an ideal laboratory for addressing
these issues.  The cluster has an age of $\sim 150-200$ Myr
\citep{1999MNRAS.306..361S,2005ApJ...622..565V} and a significantly
sub-solar metallicity [Fe/H]$\approx -0.2$
\citep{2001ApJ...549..452B}; newer spectroscopic abundance
measurements from the WIYN Open Cluster Study confirm the sub-solar
metallicity ([Fe/H]$\approx -0.14$; A.\ Steinhauer, private
communication). For the rest of this paper, we adopt [Fe/H]$=-0.2$.
The cluster main-sequence turnoff mass for the cluster is
$4.0\msun-4.6\msun$ based on the most recent Padova stellar
evolutionary models \citep{2007A&A...469..239M,2008A&A...482..883M}.
All cluster WDs therefore are the remnants of some of the most massive
stars that form WDs.

Several determinations of the cluster distance modulus and extinction
exist in the literature.  For the sake of comparison, we have adjusted
these published values to our assumed metallicity using the relation
of \citet{1998ApJ...504..170P}.  Resulting apparent distance moduli
$(m-M)_V$ include 10.3 \citep{1997AJ....114.2556T}, $10.4\pm0.1$
\citep{1999MNRAS.306..361S}, $10.42\pm 0.13$
\citep{2004MNRAS.351..649K}, and $10.26\pm 0.12$
\citep{2004AJ....127..991S}.  Published values of the reddening span
color excess values of $\ebv=0.19$ \citep{1997AJ....114.2556T} to 
$\ebv=0.255$ \citep{1999MNRAS.306..361S}.  For this paper, we 
adopt $(m-M)_V=10.3\pm 0.1$ and $\ebv=0.22\pm 0.03$ as representative
of the magnitude of and scatter in these values.

The first search for WDs in M35 was undertaken by
\citet{1980ApJ...235..992R}, who used photographic images of the
cluster to identify four WD candidates. \citet{1988A&A...202...77R}
obtained spectra of three of these stars, confirming that all three
are WDs and establishing two as likely members.  In recent years, deep
CCD imaging studies of \citet{2002AJ....124.1555V} and
\citet{2003AJ....126.1402K} identified several faint, blue stars
likely to be WDs, but these studies lack the spectroscopic
observations necessary to determine the WD masses and cooling ages.

Therefore, we began our own imaging and spectroscopic study of the WD
population of M35.  In \citet{2004ApJ...615L..49W}, we presented
initial data on eight DA (hydrogen-atmosphere) WDs in the field of NGC
2168, seven of which were claimed to be cluster members.  In this
paper, we present data for an additional six WDs, four of which are
DAs, as well as significant additional data on three of the
previously-published DAs.  All of the older data have been completely
re-reduced to correct some errors in the initial reduction, to
incorporate improvements to the atmospheric fitting routines, and to
use the newest WD and stellar evolutionary models. The fits in this
paper therefore supersede those published in
\citet{2004ApJ...615L..49W}.

\section{Cluster Photometry\label{sec.phot}}
\subsection{Observations\label{sec.phot.obs}}

Initial \ubv\ imaging of M35 was obtained on 2001 September 22 UT
using the Prime Focus Camera (PFCam) on the Lick Observatory 3m Shane
telescope.  Weather was photometric, and seeing was moderate and
steady at $\approx 1\farcs6$.  The imaging covered two $\sim
10\arcmin\times 10\arcmin$ fields, one centered $\sim 4\farcm 5$
northeast of the cluster center, and the other $\sim 4\farcm 5$
southwest of the cluster center.  As data from earlier in the night
showed evidence for a nonlinear CCD response
\citep[see][]{2007AJ....133.1490W}, these photometric data are not
presented in this paper.  However, they were of sufficient quality to
identify blue-excess objects as WD candidates for initial
spectroscopy.

Additional \ubv\ imaging of the cluster was obtained on 2004 January
23 UT with the Mosaic-1 camera on the Kitt Peak Mayall 4m telescope.
Weather was photometric and seeing was excellent: $0\farcs 75$ in $V$,
$0\farcs 9$ in $B$, and $1\farcs 4$ in $U$.  The Mosaic camera, which has a
field-of-view of $36\arcmin \times 36\arcmin$, was centered on
the cluster core.  In each filter, three long exposures were
obtained for a total of 2700s in $U$ and 360s in $B$ and $V$.  Between
each exposure, a dither of $\sim 52\arcsec$ north and east was
executed.  Single shorter exposures of 30s and 5s were also obtained in each
filter.

\subsection{Photometric reduction\label{sec.phot.redux}}
The Mosaic data were reduced using the MSCRED package of
IRAF.\footnote{IRAF is distributed by the National Optical Astronomy
Observatory, which is operated by the Association of Universities for
Research in Astronomy, Inc., under cooperative agreement with the
National Science Foundation. }  We closely followed the prescription
formulated for the NOAO Deep Wide-Field Survey (NDWFS) described in
online notes by \citet{Jannuzi2003}.\footnote{Available at \\
http://www.noao.edu/noao/noaodeep/ReductionOpt/frames.html.}  In
short, we subtracted overscans and trimmed each image, and we combined
and subtracted bias frames to remove any residual bias structure.  We
also combined and applied dome flat fields in each filter to the
science images.  We refined the image world coordinate systems using
coordinates from the Guide Star Catalog 2 and projected the images
onto the tangent plane, thereby resampling each pixel and correcting
for the variable pixel scale.  At this point in reduction we deviated
from the NDWFS reduction techniques.  In order to obtain the most
accurate photometry and to account correctly for chip-to-chip variations
in color terms \citep{2002ApJ...576..880S}, we divided each resampled
image into individual subimages for each chip; each
subimage was analyzed independently.

Object detection and photometric measurements were performed using the
DAOPHOT II suite of analysis programs \citep{1987PASP...99..191S} and
a detection threshold of $4\sigma$.  Approximately 50 bright, isolated
stars were used in each subimage to define a point-spread function
(PSF); photometry for each star was determined via
PSF-fitting. Curve-of-growth analysis of the PSF stars determined the
aperture corrections used to convert PSF magnitudes to total
instrumental magnitudes.

The photometric catalogs from each band were then matched to create a
final photometric catalog.  Stars were required to have been detected
on at least one frame in each of the three bandpasses in order to make
the final catalog.  Due to high background extinction, star-galaxy
separation was deemed unnecessary.

\subsection{Calibration\label{sec.phot.calib}}
Calibrations for the MOSAIC data were determined via imaging of
standard star fields from \citet{1992AJ....104..340L}.  The following
transformation equations were used for calibration of the data:
\\
\\
\begin{eqnarray}
u  & = & U + 2.5\log t_{\rm exp} + A_0 + A_1 (\ub) \\ & & + A_2(X-1.25) + A_3(\ub)^2 \label{eqn.u} \nonumber \\
b  & = & B + 2.5\log t_{\rm exp} + B_0 + B_1 (\bv) \\ & & + B_2(X-1.25) \label{eqn.b} \nonumber \\
v  & = & V + 2.5\log t_{\rm exp} + C_0 + C_1 (\bv) \\ & & + C_2(X-1.25) \label{eqn.v} \, , \nonumber
\end{eqnarray}
where $u,\, b,\, v$ are the total instrumental magnitudes, $U,\, B,\,
V$ are the standard system magnitudes, $t_{\rm exp}$ is the exposure
time, and $X$ is the airmass.  Examination of the standard star
observation residuals revealed the necessity of the quadratic term in
the $U$-band transformation. The number of observed standard stars
with extreme colors was fairly small, rendering the value of the
quadratic coefficient uncertain.  However, the agreement of the WD
$U$-band photometry with theoretical values gives us confidence that
this value is roughly correct; solving the transformation equation
without the quadratic term leads to observed WD colors in conflict
with the evolutionary models.  Values for the transformation
coefficients are given in Table \ref{tab.photcoeff}.

\begin{deluxetable*}{ccccc}
\tablecolumns{5}
\tablewidth{0pt}
\tabletypesize{\footnotesize}
\tablecaption{Photometric Transformation Equation Coefficients}
\tablehead{\colhead{Filter} & \colhead{Zero Point} & 
           \colhead{Color Term} & \colhead{Airmass Term} &
           \colhead{Quadratic Term} \label{tab.photcoeff}}
\startdata
$U$ & $A_0=-22.981\pm 0.009$ & $A_1 = -0.126\pm 0.021$ & $A_2 = 0.395\pm 0.059$ & $A_3 = 0.085\pm 0.019$ \\ 
$B$ & $B_0=-24.984\pm 0.006$ & $B_1 = -0.097\pm 0.005$ & $B_2 = 0.205\pm 0.031$ & \nodata \\ 
$V$ & $C_0=-25.157\pm 0.004$ & $C_1 = \phantom{-}0.041\pm 0.004$ & $C_2 = 0.078\pm 0.021$ & \nodata \\ 
\enddata
\end{deluxetable*}

As the number of observed standard stars was insufficient to determine
individual color terms for each CCD in the Mosaic camera, a single,
mean value was determined.  For purposes of comparison, a second
calibration was performed adopting published color terms for each
individual CCD \citep{Massey1999}; the resulting photometry for
individual stars in $B$ and $V$ averaged 0.01 mag fainter with a
dispersion of 0.01 mag.  As \citet{Massey1999} did not use a quadratic
term in the $U$-band transformation, direct comparison of $U$-band
terms is not possible, but \citet{Massey1999} note the chip-to-chip
color-term variation introduces a scatter of $\sim 0.04$ mag over a 1
mag range in \ub.

\begin{figure}
\includegraphics[angle=270,scale=0.33]{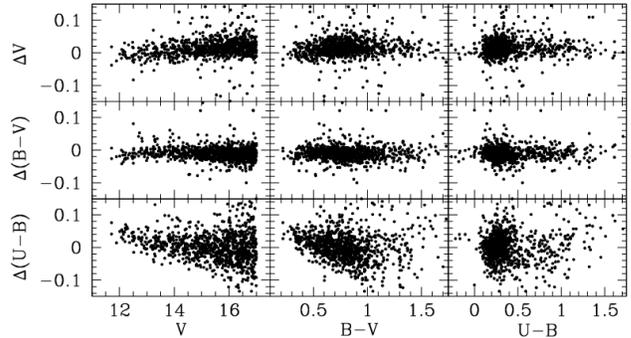}
\caption{Difference between photometry of \citet{1999MNRAS.306..361S}
and our photometry.  Only slight systematics are observed
in the $B$ and $V$-band photometry, while our \ub\  colors are
systematically bluer for hotter stars.  \label{fig.sung}}
\end{figure}

As a further sanity check, we compare our photometry with previous CCD
photometry of M35 from \citet{1999MNRAS.306..361S}, shown in Figure
\ref{fig.sung}.  We find no systematic offsets in the $B$ or $V$
photometry at $> 0.02$ mag levels.  However, a significant
systematic trend is observable in $U$: for the bluest objects, we
obtain $U$ magnitudes systematically brighter by $\approx 0.03$ mag.
It is unclear if this systematic is intrinsic to our data, perhaps
indicative that the color terms in Eq.~\ref{eqn.u} are a poor
approximation to the true color response \citep[see discussions of
\ub\ photometry in][]{1995PASP..107..672B}, if the systematic is
intrinsic to the \citet{1999MNRAS.306..361S} data, or both.

\subsection{Checking the Cluster Distance, Reddening  and Metallicity}

\begin{figure}
\plotone{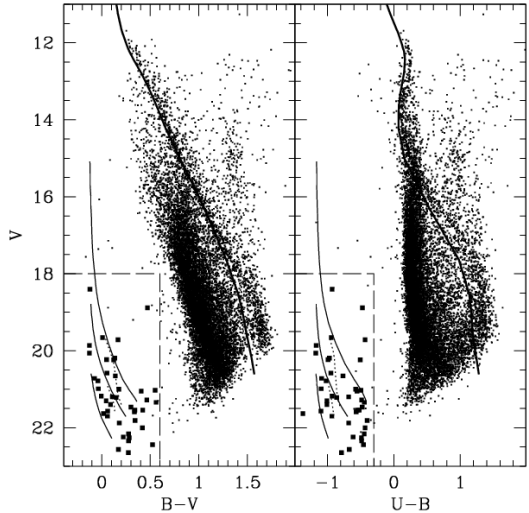}
\caption{Color-magnitude diagrams for the field of M35.  The thick,
  solid line is a 175-Myr, $Z=0.013$ Padova isochrone.
  White dwarf cooling models are plotted for
  DA (solid) and DB (dotted) WD models with masses 0.4\msun, 0.8\msun,
  and 1.2\msun\ (brighter to fainter) for cooling ages $\leq 200$Myr;
  DB models are only plotted for $\teff \leq 30000$K. The 1.2\msun\ DB
  model is omitted, as it appears as a single point nearly
  coincident with the faint end of the 1.2\msun\ DA model.  All models
  are shifted to the cluster distance and reddening.  White dwarf
  photometric selection criteria are shown as long-dashed lines;
  selected white dwarf candidates are indicated by squares. Error bars
  in white dwarf photometry are smaller than the point size.
  \label{fig.init_cmd}}
\end{figure}

The $\bv,\, V$ and $\ub,\, V$ color-magnitude diagrams (CMDs) of the
stars in the field of M35 are shown in Figure \ref{fig.init_cmd}.
Overplotted is the Padova isochrone for a $Z=0.013$
\citep[$\mathrm{[Fe/H]}=-0.2$ for the traditional solar abundances
of][]{1989GeCoA..53..197A}, 200-Myr-old stellar population shifted to
the adopted cluster distance and reddening (see
\S\ref{sec.intro.m35}).  The isochrone is found to be an excellent
representation of the main sequence for $V\lesssim 17$ ($M_V\lesssim
6.5$) in the \bv\ CMD.  

The same $Z=0.013$, 200-Myr isochrone is too blue in the \ub\ CMD;
agreement for $V\lesssim 16$ ($M_V\lesssim 5.5$) can be reached by
increasing the reddening to $\ebv=0.3$. Such a change in reddening
throws off the observed agreement in the \bv\ CMD.  This suggests
either a non-standard reddening law in the direction of M35, a
significant error in our $U$-band zeropoint (in a direction which
would worsen the systematic offset mentioned in the previous section),
or a problem with the isochrone $U$-band calculations.  Additional
data and analysis are needed to explore this discrepancy further.

A comparison of our data with $Z=0.008$ and $Z=0.019$ Padova
isochrones finds that the observed main sequence agrees with the
isochrone if the distance modulus of the cluster is shifted 0.15 mag
closer or further, respectively.  As with the $Z=0.013$ isochrone, the
\ub\ CMD requires a higher assumed reddening value to bring the
isochrone and main sequence into agreement.

The observed main sequence is also in excellent agreement with the
empirical zero-age main sequence of \citet{1981A&A....97..235M},
shifted to the adopted cluster distance and reddening, and with the
fiducial M35 main sequence of \citet{2002AJ....124.1555V}.

We therefore conclude that our photometry is consistent with the
literature-based distance, reddening, and metallicity we adopted in
\S\ref{sec.intro.m35}. We emphasize that, as our photometric data saturate
well below the main sequence turnoff, our main sequence data are
not useful for age determination.

\subsection{Candidate WD Selection\label{sec.phot.cands}}

\begin{figure}
\plotone{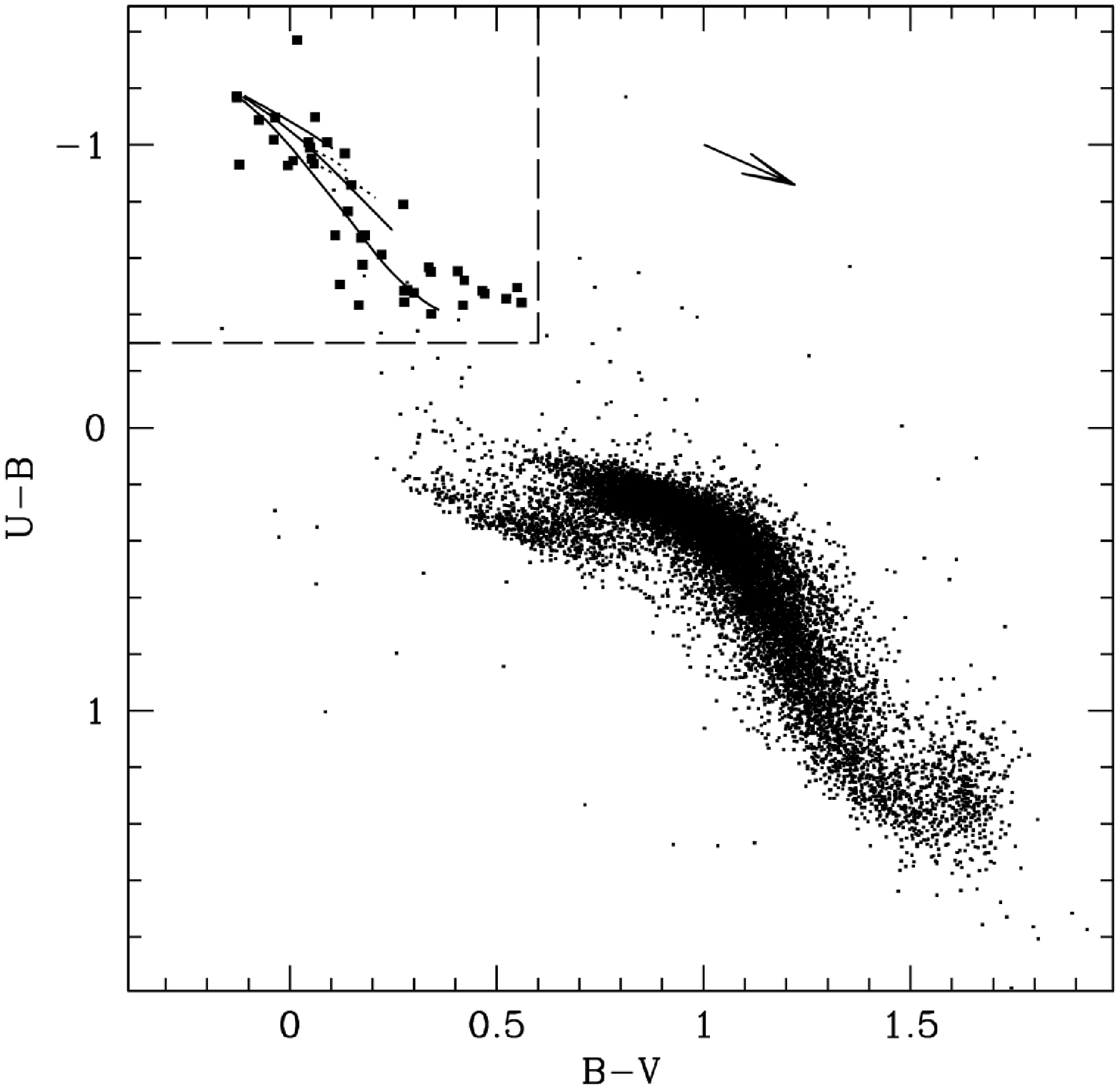}
\caption{Color-color plot for the field of M35.  Small points are for
  all detected objects; larger squares are the selected WD candidates.
  Cooling models are shown for $\tcool\leq 200$Myr for DA WDs (solid)
  of masses 0.4\msun, 0.8\msun, and 1.2\msun\ (bottom to top) and for
  DB WDs of masses 0.4\msun\ and 0.8\msun; the 1.2\msun\ model is
  omitted. Models are shifted to the assumed cluster reddening.  The
  dashed line indicates WD selection criteria. The arrow indicates the
  reddening vector of $\ebv=0.22$, assuming $R_V=3.1$.
  \label{fig.init_cc}}
\end{figure}

The CMDs spanning the entire magnitude range of our photometry are
shown in Figure \ref{fig.init_cmd} along with cooling curves for
hydrogen-atmosphere (DA) and helium-atmosphere (DB) WDs for a range of
WD masses, all shifted to the adopted cluster distance modulus and
reddening.  Several faint, blue objects are seen in the region of the
CMD that should be populated by cluster WDs.  The $\ub,\,\bv$
color-color plot (Figure \ref{fig.init_cc}) more clearly shows those
objects with colors consistent with WD cooling models.

Candidate WDs are selected using the criteria $V\geq 18$, $\bv\leq
0.6$, and $\ub\leq -0.3$.  These criteria would include all WDs at the
cluster distance and reddening except for the hottest low-mass
($M_{\rm WD}\lesssim 0.4\msun, \, \teff\gtrsim 50,000$K) WDs.  These
criteria are indicated in Figures\ \ref{fig.init_cmd} and
\ref{fig.init_cc}; candidate WDs selected via these criteria indicated
as larger solid points in the CMDs and color-color plot.  Coordinates,
photometry, and cross-identifications for each candidate WD are given
in Table \ref{tab.cands}.

\begin{deluxetable*}{lccccccccl}
\tabletypesize{\footnotesize}
\tablecolumns{10}
\tablewidth{0pt}
\tablecaption{Candidate White Dwarfs in NGC 2168\label{tab.cands}}
\tablehead{\colhead{Object} & \colhead{RA} & \colhead{Dec} & \colhead{$V$} &
     \colhead{$\delta V$} & \colhead{$B-V$} & \colhead{$\delta(B-V)$} & 
     \colhead{$\ub$} & \colhead{$\delta(\ub)$} & 
     \colhead{Cross-Identification}}
\startdata
LAWDS  1 & 6:08:38.79 & 24:15:06.9 & 20.989 &  0.019 &$-0.035$&  0.028 &$-1.095$&  0.028 & NGC 2168-1 in \citet{1980ApJ...235..992R} \\ 
LAWDS  2 & 6:08:42.30 & 24:10:17.7 & 21.569 &  0.032 & \phantom{$-$}0.061 &  0.044 &$-1.097$&  0.042 & \\ 
LAWDS  3 & 6:09:04.78 & 24:21:39.2 & 20.581 &  0.020 & \phantom{$-$}0.053 &  0.028 &$-0.950$&  0.028 & \\ 
LAWDS  4 & 6:09:05.76 & 24:12:11.8 & 21.216 &  0.021 & \phantom{$-$}0.148 &  0.030 &$-0.858$&  0.030 & NGC 2168-2 in \citet{1980ApJ...235..992R} \\ 
LAWDS  5 & 6:09:11.54 & 24:27:20.9 & 20.065 &  0.017 &$-0.128$&  0.024 &$-1.173$&  0.025 & NGC 2168-3 in \citet{1980ApJ...235..992R} \\ 
LAWDS  6 & 6:09:23.48 & 24:27:22.0 & 19.863 &  0.016 &$-0.128$&  0.023 &$-1.167$&  0.024 & NGC 2168-4 in \citet{1980ApJ...235..992R} \\ 
LAWDS 10 & 6:09:43.63 & 24:19:15.8 & 20.238 &  0.020 & \phantom{$-$}0.122 &  0.030 &$-0.506$&  0.032 & \\ 
LAWDS 11 & 6:09:42.79 & 24:11:05.4 & 21.198 &  0.025 & \phantom{$-$}0.110 &  0.037 &$-0.680$&  0.038 & \\ 
LAWDS 12 & 6:09:31.19 & 24:19:06.2 & 21.303 &  0.026 & \phantom{$-$}0.045 &  0.038 &$-1.009$&  0.038 & \\ 
LAWDS 13 & 6:09:29.71 & 24:15:58.6 & 21.542 &  0.024 & \phantom{$-$}0.341 &  0.037 &$-0.551$&  0.038 & \\ 
LAWDS 14 & 6:09:15.10 & 24:33:15.4 & 21.701 &  0.027 & \phantom{$-$}0.059 &  0.038 &$-0.933$&  0.036 & \\ 
LAWDS 15 & 6:09:11.63 & 24:02:38.5 & 20.785 &  0.022 &$-0.039$&  0.032 &$-1.019$&  0.033 & \\ 
LAWDS 18 & 6:08:35.44 & 24:34:19.5 & 22.566 &  0.043 & \phantom{$-$}0.172 &  0.071 &$-0.671$&  0.069 & \\ 
LAWDS 22 & 6:08:24.65 & 24:33:47.6 & 19.657 &  0.016 & \phantom{$-$}0.008 &  0.023 &$-0.944$&  0.025 & \\ 
LAWDS 27 & 6:09:06.26 & 24:19:25.3 & 21.398 &  0.026 & \phantom{$-$}0.090 &  0.039 &$-1.009$&  0.039 & Discovered by \citet{2002AJ....124.1555V}\\ 
LAWDS 28 & 6:08:13.50 & 24:20:32.5 & 21.631 &  0.029 & \phantom{$-$}0.018 &  0.042 &$-1.370$&  0.040 & \\ 
LAWDS 29 & 6:08:02.20 & 24:25:24.2 & 20.719 &  0.019 &$-0.075$&  0.027 &$-1.087$&  0.027 & \\ 
LAWDS 30 & 6:07:56.63 & 24:13:27.2 & 21.175 &  0.020 &$-0.005$&  0.029 &$-0.927$&  0.030 & \\ 
LAWDS 31 & 6:10:08.48 & 24:22:32.3 & 20.231 &  0.020 & \phantom{$-$}0.048 &  0.031 &$-0.991$&  0.033 & \\ 
LAWDS 32 & 6:09:37.35 & 24:31:52.5 & 22.000 &  0.030 & \phantom{$-$}0.418 &  0.047 &$-0.432$&  0.062 & \\ 
LAWDS 33 & 6:09:33.02 & 24:15:23.5 & 22.259 &  0.033 & \phantom{$-$}0.284 &  0.051 &$-0.488$&  0.060 & \\ 
LAWDS 34 & 6:09:25.26 & 24:14:05.2 & 21.026 &  0.022 & \phantom{$-$}0.550 &  0.033 &$-0.495$&  0.035 & \\ 
LAWDS 35 & 6:09:12.92 & 24:22:17.4 & 21.809 &  0.027 & \phantom{$-$}0.342 &  0.040 &$-0.402$&  0.042 & \\ 
LAWDS 36 & 6:09:04.59 & 24:06:45.7 & 21.276 &  0.021 & \phantom{$-$}0.465 &  0.033 &$-0.483$&  0.036 & \\ 
LAWDS 37 & 6:08:59.02 & 24:08:40.8 & 21.049 &  0.020 & \phantom{$-$}0.406 &  0.030 &$-0.554$&  0.033 & \\ 
LAWDS 38 & 6:09:08.26 & 24:36:24.2 & 22.158 &  0.034 & \phantom{$-$}0.277 &  0.051 &$-0.444$&  0.056 & \\ 
LAWDS 39 & 6:09:01.46 & 24:26:50.3 & 22.648 &  0.052 & \phantom{$-$}0.275 &  0.081 &$-0.790$&  0.086 & \\ 
LAWDS 40 & 6:08:54.53 & 24:35:58.5 & 22.244 &  0.043 & \phantom{$-$}0.222 &  0.061 &$-0.612$&  0.070 & \\ 
LAWDS 41 & 6:08:34.97 & 24:32:47.3 & 18.886 &  0.015 & \phantom{$-$}0.471 &  0.022 &$-0.473$&  0.024 & \\ 
LAWDS 42 & 6:08:24.11 & 24:22:34.7 & 22.338 &  0.042 & \phantom{$-$}0.277 &  0.069 &$-0.484$&  0.076 & \\ 
LAWDS 43 & 6:08:20.96 & 24:08:51.4 & 22.440 &  0.037 & \phantom{$-$}0.523 &  0.065 &$-0.456$&  0.086 & \\ 
LAWDS 44 & 6:08:03.84 & 24:27:37.2 & 19.718 &  0.017 & \phantom{$-$}0.167 &  0.025 &$-0.433$&  0.027 & \\ 
LAWDS 45 & 6:08:03.77 & 24:27:38.3 & 18.403 &  0.015 &$-0.122$&  0.021 &$-0.929$&  0.022 & \\ 
LAWDS 46 & 6:08:00.63 & 24:07:40.2 & 20.998 &  0.019 & \phantom{$-$}0.176 &  0.029 &$-0.575$&  0.030 & \\ 
LAWDS 47 & 6:08:00.37 & 24:18:02.1 & 21.540 &  0.023 & \phantom{$-$}0.422 &  0.036 &$-0.521$&  0.038 & \\ 
LAWDS 48 & 6:07:58.70 & 24:28:40.1 & 21.590 &  0.030 & \phantom{$-$}0.336 &  0.047 &$-0.567$&  0.048 & \\ 
LAWDS 49 & 6:07:52.95 & 24:25:23.5 & 20.191 &  0.017 & \phantom{$-$}0.133 &  0.025 &$-0.970$&  0.026 & \\ 
LAWDS 50 & 6:07:47.77 & 24:28:52.0 & 21.465 &  0.030 & \phantom{$-$}0.300 &  0.044 &$-0.477$&  0.046 & \\ 
LAWDS 51 & 6:07:47.39 & 24:34:15.3 & 21.877 &  0.031 & \phantom{$-$}0.182 &  0.044 &$-0.680$&  0.041 & \\ 
LAWDS 52 & 6:07:47.37 & 24:35:23.5 & 20.653 &  0.018 & \phantom{$-$}0.140 &  0.027 &$-0.764$&  0.028 & \\ 
LAWDS 53 & 6:07:33.79 & 24:07:55.5 & 21.341 &  0.028 & \phantom{$-$}0.560 &  0.046 &$-0.442$&  0.050 &
V57 in \citet{2004AJ....128..312M}\\ 
\enddata
\tablecomments{Units of right ascension are hours, minutes and seconds, and 
units of declination are degrees, arcminutes, and arcseconds.  Coordinates 
are for Equinox J2000.0.  The official (i.e., IAU-approved) format for
each object name is NGC 2168:LAWDS NN.}
\end{deluxetable*}

\begin{figure*}
\begin{center}
\includegraphics[scale=0.67]{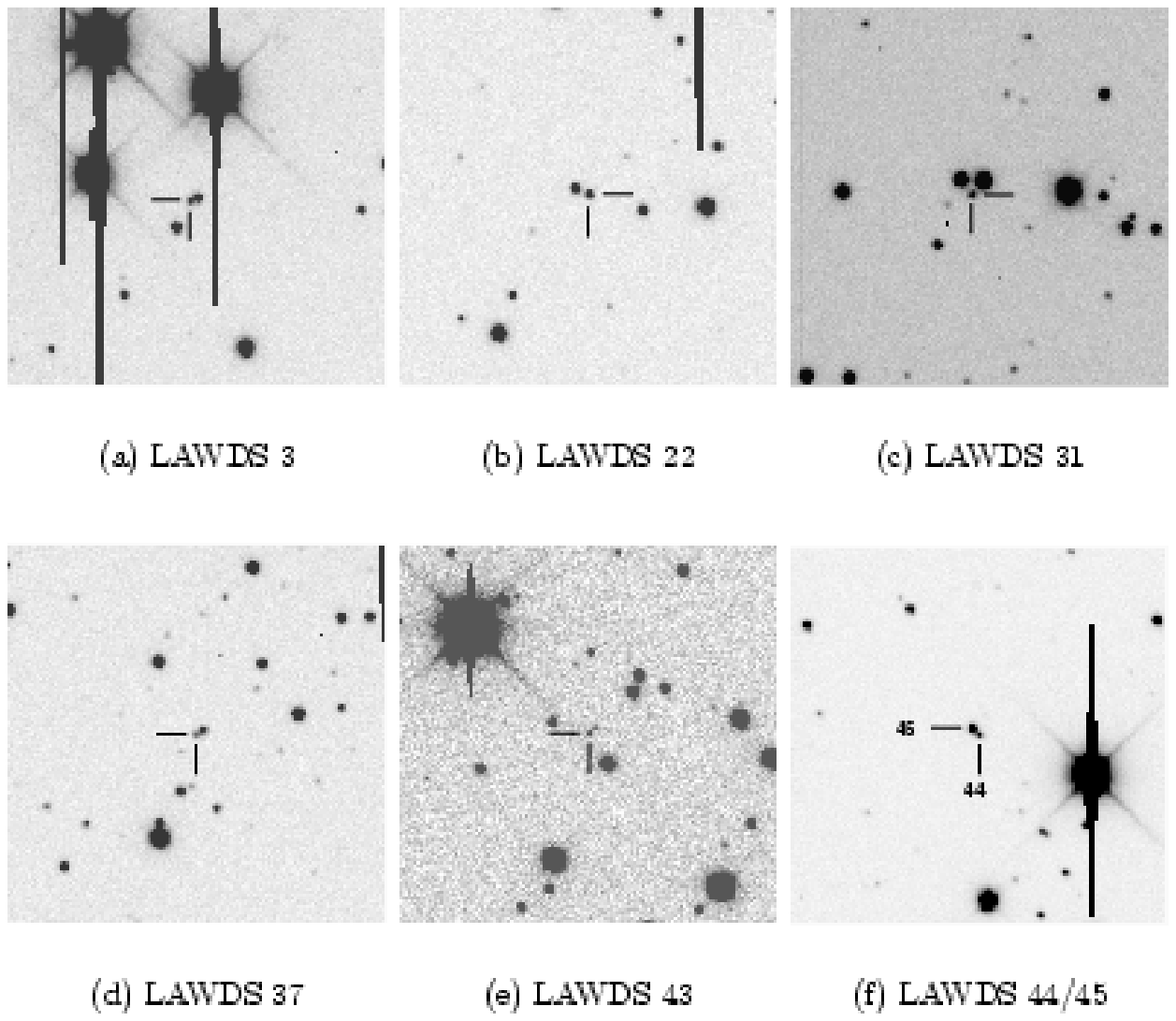}
\end{center}
\caption{Finder charts for white dwarf candidates with close optical
  companions.  Each image is 1\arcmin\ on a side.  North is to the
  left; east is down. Note that in panel (f), each component in an
  optical binary is identified. \label{fig.bin_find}}
\end{figure*}

While most of the WD candidates are fairly isolated, some
have close optical companions.  In order to clarify identification for
these WD candidates, finder charts for objects with close companions 
are provided in Figure \ref{fig.bin_find}.

\section{WD Spectroscopy\label{sec.spec}}
\subsection{Observations and Data Reduction\label{sec.spec.obs}}
Spectra of several WD candidates were obtained between 2002 December
and 2005 November with the blue camera of the LRIS spectrograph on
Keck I \citep{1995PASP..107..375O,1998SPIE.3355...81M}. We selected
the 400 grooves mm$^{-1}$, 3400\AA\ blaze grism for the highest
available throughput for the vital higher-order Balmer lines. The D560
dichroic was used to obtain simultaneous observations of the H$\alpha$
line (not presented here). Our aperture was a 1\arcsec\ longslit at
parallactic angle. The resulting spectral resolution (FWHM) is
$\approx 6$\AA.

We reduced the spectra using the \emph{onedspec} package in
IRAF. Overscan regions were used to subtract the amplifier bias. Flat
fielding was complicated by the discovery of a sharp inflection point
in the response at $\approx 4200$\AA\ and two low-level ($\approx
3\%$) emission lines in the illuminated dome flat field spectra
between 3950\AA\ and 3975\AA; both of these features introduced
ringing into the flat field.  The ringing due to the inflection point
was eliminated by creating a piecewise-smooth response function for
image sections below and above the inflection point. As internal flat
fields lacked the emission features, only internal flats were used to
create the final flat fields.

 Cosmic rays were removed from the two-dimensional spectrum using the
``L.A.\ Cosmic'' Laplacian cosmic-ray rejection routine
\citep{2001PASP..113.1420V}. We then averaged multiple exposures of
individual objects weighted by their individual signal-to-noise
ratios and extracted the one-dimensional spectrum. We
applied a wavelength solution derived from Hg, Cd, and Zn arclamp
spectra. We determined and applied a relative flux calibration from
long-slit spectra of multiple spectrophotometric standard stars. We
made no attempt at obtaining absolute spectrophotometry for any
object.

\begin{deluxetable}{llccc}
\tabletypesize{\footnotesize}
\tablecolumns{6}
\tablewidth{0pt}
\tablecaption{Spectroscopically-Observed Objects\label{tab.wdspec}}
\tablehead{\colhead{Object} & \colhead{UT Obs Date} & \colhead{Total exp. (s)} & \colhead{S/N\tablenotemark{a}} & \colhead{ID} }
\startdata
LAWDS 1   & 2002 Dec 8  & 2700 & 64 & DA \\
LAWDS 2   & 2002 Dec 8  & 2700 &109 & DA \\
          & 2004 Oct 11 & 3600 &    &    \\
LAWDS 4   & 2002 Dec 8  & 2700 & 32 & DB \\
          & 2004 Oct 11 & 900  &    &    \\
LAWDS 5   & 2002 Dec 8  & 2700 &215 & DA \\
          & 2002 Dec 9  & 3600 &    &    \\
LAWDS 6   & 2002 Dec 8  & 2700 &247 & DA \\
          & 2002 Dec 9  & 2700 &    &    \\
LAWDS 11  & 2004 Feb 12 & 3600 & 57 & DA \\
          & 2005 Nov 26 & 1800 &    &    \\
          & 2005 Nov 27 & 4800 &    &    \\
LAWDS 12  & 2005 Nov 26 & 4800 & 91 & DA \\
LAWDS 14  & 2005 Nov 26 & 6000 & 73 & DA \\
LAWDS 15  & 2004 Feb 12 & 3300 & 86 & DA \\
LAWDS 22  & 2004 Feb 12 & 3300 &198 & DA(H?) \\
LAWDS 27  & 2002 Dec 9  & 3600 & 96 & DA \\
          & 2004 Feb 12 & 3600 &    &    \\
          & 2004 Oct 11 & 3300 &    &    \\
LAWDS 28  & 2004 Oct 12 & 1200 & 73 & DQ \\
          & 2005 Nov 26 & 4800 &    &    \\
LAWDS 29  & 2004 Oct 12 & 3600 & 75 & DA \\
LAWDS 30  & 2005 Nov 27 & 6000 & 41 & DA \\
\enddata
\tablenotetext{a}{Average of signal-to-noise per resolution element at
  pseudocontinuum surrounding H$\delta$ (for DA WDs) or at 4200\AA\ 
  (non-DAs)}
\end{deluxetable}

\begin{figure}
\includegraphics[scale=0.44]{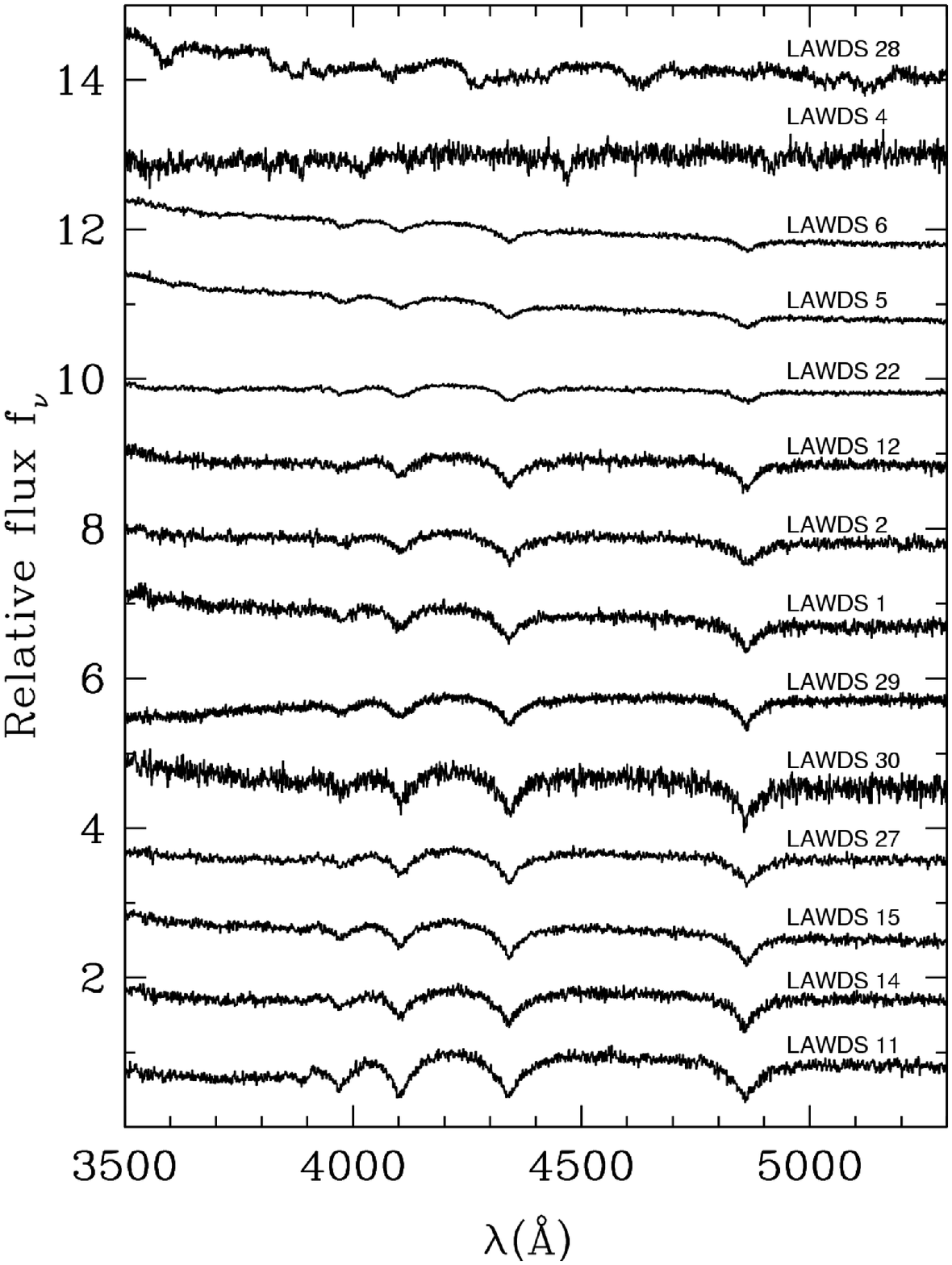}
\caption{Spectra of observed WDs in the field of NGC 2168.  Spectra
  have been normalized at 5200\AA\ and arbitrary vertical offsets
  applied.  The top two spectra are hot DQ (LAWDS 28) and the DB
  (LAWDS 4); the remaining are DAs, ordered from top to bottom in
  decreasing \teff. \label{fig.spec}}
\end{figure}

The observed candidates, their dates of observation, and the
spectroscopic identification of each object are given in Table
\ref{tab.wdspec}; all spectra are shown in Figure \ref{fig.spec}. 
NGC 2168:LAWDS 28 is the hot DQ WD discussed in
\citet{2006ApJ...643L.127W}; NGC 2168:LAWDS 4 is spectral
type DB and is discussed in \S\ref{sec.discuss.db}.  All twelve other
WDs are of spectral type DA.

\begin{figure}
\plotone{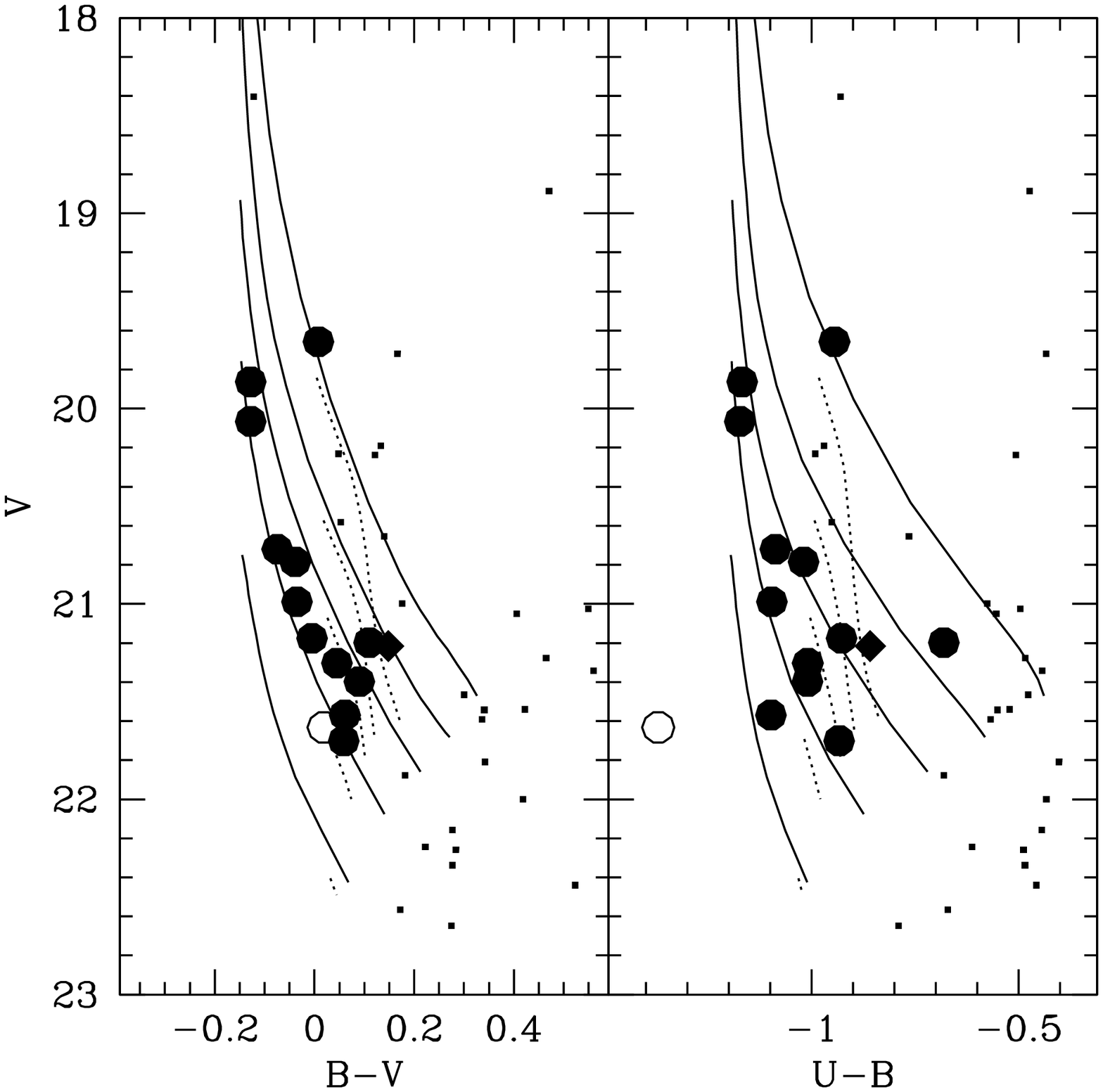}
\caption{The white dwarf regions of the M35 \bv\ (left) and \ub\
  (right) CMDs with spectroscopic identifications.  Large filled dots
  are DA white dwarfs; the open circle is the hot DQ white dwarf, and
  the DB white dwarf is plotted as a diamond.  Small points are white
  dwarf candidates not observed spectroscopically.  Curves are cooling
  models for DA (solid) and DB (dashed) white dwarfs with cooling ages
  $\leq 200$Myr for masses of 0.4\msun, 0.6\msun, 0.8\msun, 1.0\msun,
  and 1.2\msun\ WDs (top to bottom).  Curves are shifted to the
  white-dwarf derived distance and reddening values (see
  \S\ref{sec.discuss.dmod}).  In the \bv\ CMD, the cluster WDs form a
  tight cooling sequence near the $1\msun$ cooling track, though these
  are more scattered in the \ub\ CMD.\label{fig.id_cmd}}
\end{figure}

Figure \ref{fig.id_cmd} shows the WD regions of the CMDs and
color-color plot of M35 indicating each object's spectroscopic
identification.  In the \bv\ vs.\ $V$ CMD, the cluster WDs form a very
tight sequence about the 1\msun\ DA cooling model, with the exception
of the likely magnetic DA LAWDS 22 (see next section). The redder
color of LAWDS 22 could be due to an unresolved low-mass
companion; the red side of LRIS was not functioning on the night this
WD was observed, so we are unable to examine the spectra for evidence
of such a companion.  The hot DQ LAWDS 28 is significantly bluer
in \ub\ than all the other WDs.  This exceptional color is due to high carbon opacity
at wavelengths $\lesssim 1500$\AA, which re-distributes the
ultraviolet flux into the near-UV \citep{Dufour2008}.

\subsection{WD parameter determination\label{sec.spec.params}}

\teff\ and \logg\ were determined for each DA WD via simultaneous
fitting of the H$\beta$ to H9 Balmer line profiles
\citep{1992ApJ...394..228B}. Model atmospheres used for this fitting
are derived from Koester's model grids; details of the input physics
and methods can be found in \citet{2001A&A...378..556K} and references
therein.  The algorithm for our fitting routine and error estimation
are detailed in \citet{2007AJ....133.1490W}, with the change that the
model grid was expanded to include higher surface gravities
($7.0\leq\logg\leq 10.0$). Signal-to-noise was calculated by
determining the root mean square scatter per 6\AA\ resolution element
about the fit pseudo-continuum on both sides of the H$\delta$ line.

Initial spectral fits proved unsatisfactory for the hottest white
dwarfs (LAWDS 5, LAWDS 6, and LAWDS 22), with the observed spectrum
deviating strongly from the best-fit models.  The models severely
underestimated the depth of the H$\epsilon$ line and overestimated
those of lower-order Balmer lines.  There are several potential
sources for the poor H$\epsilon$ fits in these three WDs:

\begin{itemize}
\item \emph{Magnetic fields} --- In the spectrum of LAWDS 22, the
  cores of the Balmer lines are significantly flattened.  This is
  likely to be due to a magnetic field with a strength sufficient to
  broaden the line without producing obvious Zeeman splitting at our
  spectral resolution \citep[$\sim 500$ kG; ][]{2000PASP..112..873W};
  a rough comparison to a magnetic atmospheric models reveals good
  agreement for a magnetic field strength of $\sim 1$ MG (A.\ Kanaan,
  private communication). The Balmer line models we fit do not include
  magnetic fields, and so the parameters derived for this star are
  invalid.  Weaker fields could also be present in LAWDS 5 and LAWDS
  6, accounting for their poor fits.  However, there is no compelling
  evidence for magnetism in these two stars.

\item \emph{Known high \teff\  systematics} --- DA WDs with
  $\teff\gtrsim 40000$ K show systematic differences in \teff\  
  derived from different methods, though the proposed causes of these
  differences, including non-LTE effects, small quantities of helium,
  and metal opacities, remain a matter of debate
  \citep[e.g.,][]{1994ApJ...432..305B,1997A&A...322..256N,2003MNRAS.344..562B,2007ApJ...667.1126L}. The sense and magnitude of our deviations are
  similar to those calculated by \citet{1998MNRAS.299..520B} for a
  metal-rich, non-LTE atmosphere compared to a pure-H, LTE model. 

\item \emph{Interstellar absorption} --- In our highest
  signal-to-noise spectra, we observe an additional absorption feature
  on the red wing of the H$\gamma$ line that we identify as the
  4430\AA\ diffuse interstellar band \citep[DIB;
  e.g.,][]{1966ZA.....64..512H}.  Weak \ion{Ca}{2} K absorption is
  also visible in the higher signal-to-noise WD spectra; due to the
  high \teff\ of these WDs, this absorption must be interstellar.  The
  H$\epsilon$ lines will therefore be contaminated by \ion{Ca}{2} H
  absorption. In the hottest WDs, the depth of \ion{Ca}{2} lines are
  $\sim 50\%$ of the H$\epsilon$ line.  We have not attempted to model
  and subtract the \ion{Ca}{2} H line, so this line will have some
  effect on the model atmosphere fits.  However, as the \ion{Ca}{2} H
  line is significantly narrower than the H$\epsilon$ line, and the
  deviations span the entire H$\epsilon$ line, contamination from
  \ion{Ca}{2} H is therefore not the primary cause of the poor fits.

\item \emph{Shortcomings in the model atmospheres} --- The high-order
 Balmer lines are strongly affected by the dissolution of the higher
 atomic energy levels due to interactions with perturbing
 particles. This is described in the atmospheric models by the
 Hummer-Mihalas-Deppen occupation probability, which has free
 parameters that are difficult to quantify. The perturbation effect
 may be a bit too strong, leading to weaker higher-order lines in the
 models.  To test this possibility, we have used our models and
 fitting routine to determine parameters of hot, high-gravity WDs in
 the Palomar Green WD sample \citep{2005ApJS..156...47L}, whose
 spectra were graciously provided by J.\ Liebert.  These fits do not
 show the same problems with the fitting of H$\epsilon$ observed
 here; therefore, we conclude that shortcomings in the models are
 likely not responsible for our poor spectral fits.

\item \emph{Data reduction systematics} ---  As mentioned in
  \S\ref{sec.spec.obs}, the dome spectroscopic flat field lamps were
  observed to have weak emission features in this wavelength region,
  so only internal flat field lamps were used.  If the internal lamps
  have similar spectral features, the
  H$\epsilon$ profiles could be affected.  However, no such features
  were apparent in the internal flats, and the cooler DAs in this
  study do not show systematics in the H$\epsilon$ line, so we
  consider this explanation unlikely.

\end{itemize}
In short, we find no compelling explanation why the H$\epsilon$ lines in
LAWDS 5 and LAWDS 6 are not well-represented by the atmospheric
models; the failure in LAWDS 22 seems likely to be due to magnetic
fields. 

As a check against model atmosphere or data reduction systematics
affecting all of our WDs, we re-fit each WD excluding the H$\epsilon$
line from the fits.  In all cases, the \teff\ values are identical
within the internal fitting errors (described below).  This is
expected, as the majority of leverage in \teff\ determination is from
the lower Balmer lines. The \logg\ values are systematically higher by
an average of 0.1 dex when the H$\epsilon$ line is excluded. This
change is because the higher-order Balmer lines are most sensitive to
surface gravity, and the absence of the H8 and H9 lines due to the
dissolution of these energy levels in these higher gravity atmospheres
only sets a lower limit on the surface gravity.

\begin{deluxetable*}{lcccccccccccccc}
\tabletypesize{\scriptsize}
\tablecolumns{15}
\tablewidth{0pt}
\tablecaption{DA White Dwarf Spectral Fits and Derived Parameters\label{tab.wdfits}}
\tablehead{ \colhead{Object} & \colhead{\teff} & \colhead{\logg} & \multicolumn{2}{c}{Internal Errors}
  & \multicolumn{2}{c}{Total Errors} & \colhead{$\mwd$} & \colhead{$\delta \mwd$} 
  & \colhead{\tcool} & \colhead{$\delta\tcool$} & \colhead{$\minit$} 
  & \colhead{$\delta M_{\rm init, Obs}$\tablenotemark{a}} & \colhead{$\delta M_{\rm init, Sys}$\tablenotemark{b}} \\
  & \colhead{(K)} & & \colhead{$\delta\teff$} & \colhead{$\delta\logg$} 
  & \colhead{$\delta\teff$} & \colhead{$\delta\logg$} & \colhead{(\msun)} & \colhead{(\msun)} 
  & \colhead{$\log ({\rm yr})$} & \colhead{$\log ({\rm yr})$} & \colhead{(\msun)} 
  & \colhead{(\msun)} & \colhead{(\msun)}}
\startdata
LAWDS 1\tablenotemark{c}  & 33400 & 8.36 & 320 & 0.11 & 1150 & 0.16 & 0.873 & 0.091 & 7.228 & 0.260 & 4.39 & $^{+0.23}_{-0.09}$ & $^{+0.35}_{-0.27}$\\
LAWDS 2  & 34100 & 8.62 & 310 & 0.04 & 1140 & 0.13 & 1.015 & 0.067 & 7.657 & 0.202 & 4.79 & $^{+0.47}_{-0.26}$ & $^{+0.46}_{-0.36}$\\
LAWDS 5\tablenotemark{c}  & 53750\tablenotemark{d} & 8.39\tablenotemark{d} & 850 & 0.05 & 1390 & 0.13 & 0.916 & 0.075 & 6.103 & 0.088 & 4.21 & $^{+0.00}_{-0.00}$ & $^{+0.29}_{-0.23}$\\
LAWDS 6\tablenotemark{c}  & 56000\tablenotemark{e} & 8.32\tablenotemark{e} &1500 & 0.14 & 1860 & 0.18 & 0.877 & 0.096 & 6.077 & 0.082 & 4.21 & $^{+0.00}_{-0.00}$ & $^{+0.29}_{-0.23}$\\
LAWDS 11 & 20800 & 8.28 & 520 & 0.08 & 1260 & 0.16 & 0.802 & 0.096 & 8.047 & 0.173 & 6.63 & $^{+0.00}_{-1.38}$ & $^{+1.89}_{-0.96}$\\
LAWDS 12 & 34500\tablenotemark{d} & 8.44\tablenotemark{d} & 280 & 0.10 & 1140 & 0.16 & 0.922 & 0.092 & 7.314 & 0.281 & 4.43 & $^{+0.30}_{-0.13}$ & $^{+0.36}_{-0.27}$\\
LAWDS 14 & 29200 & 8.62 & 220 & 0.07 & 1120 & 0.14 & 1.010 & 0.072 & 7.865 & 0.167 & 5.32 & $^{+0.95}_{-0.44}$ & $^{+0.73}_{-0.48}$\\
LAWDS 15\tablenotemark{c} & 29500 & 8.40 & 160 & 0.00 & 1110 & 0.15 & 0.888 & 0.088 & 7.551 & 0.236 & 4.63 & $^{+0.38}_{-0.22}$ & $^{+0.41}_{-0.32}$\\
LAWDS 22\tablenotemark{c} & 50000\tablenotemark{d} & 8.08\tablenotemark{d} &1420 & 0.07 & 1800 & 0.12 & \nodata\tablenotemark{f} & \nodata\tablenotemark{f} & \nodata\tablenotemark{f} & \nodata\tablenotemark{f} & \nodata\tablenotemark{f} & \nodata\tablenotemark{f} & \nodata\tablenotemark{f} \\
LAWDS 27 & 30200 & 8.64 & 100 & 0.04 & 1100 & 0.13 & 1.022 & 0.072 & 7.840 & 0.170 & 5.22 & $^{+0.71}_{-0.42}$ & $^{+0.68}_{-0.45}$\\
LAWDS 29 & 32400 & 8.38 & 220 & 0.07 & 1120 & 0.14 & 0.882 & 0.078 & 7.331 & 0.235 & 4.44 & $^{+0.23}_{-0.12}$ & $^{+0.36}_{-0.28}$\\
LAWDS 30 & 30400 & 8.62 & 390 & 0.18 & 1170 & 0.22 & 1.011 & 0.122 & 7.808 & 0.289 & 5.12 & $^{+1.49}_{-0.55}$ & $^{+0.63}_{-0.43}$\\
\enddata
\tablenotetext{a}{Error due to total fitting error}
\tablenotetext{b}{Additional systematic error due to cluster age
  uncertainty}
\tablenotetext{c}{Re-reduction of data presented in \citet{2004ApJ...615L..49W}}
\tablenotetext{d}{Fit excludes H$\epsilon$}
\tablenotetext{e}{Adopted values intermediate to fits obtained by including and excluding H$\epsilon$}
\tablenotetext{f}{Values not calculated due to poor Balmer line fits}
\end{deluxetable*}
 
\begin{figure}
\includegraphics[scale=0.44]{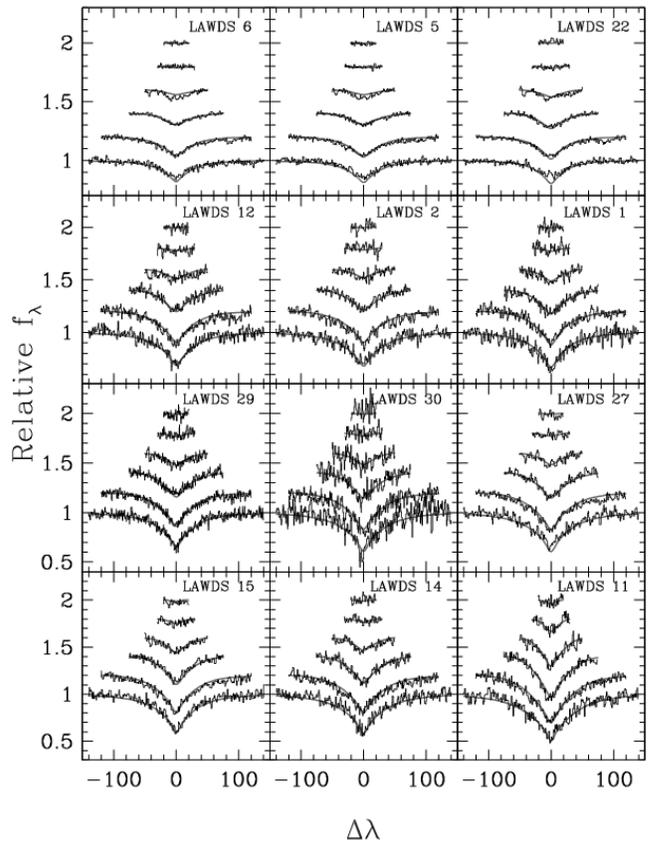}
\caption{Balmer line profiles and best-fit models for 12 DA white
  dwarfs in the field of M35, from H$\beta$ (bottom) to H9
  (top). Plotted models are the adopted best-fit models from Table
  \ref{tab.wdfits}.  Significant deviations in H$\epsilon$ from the
  best-fit models are noticeable in LAWDS 6, LAWDS 5, and LAWDS 22.
  The cores of the lines for LAWDS 22 are noticeable flattened, likely
  indicative of nearly-resolved Zeeman splitting.  The 4430\AA\
  diffuse interstellar band can be seen in the red (right) wing of
  H$\gamma$ in many of the higher signal-to-noise WDs, and
  interstellar \ion{Ca}{2} K can be often seen in the blue wing of
  H$\epsilon$. WDs are arranged in order of decreasing
  \teff. \label{fig.wdfits}}
\end{figure}

The best-fitting model atmospheres from each of the two fits were
plotted over each WD's observed spectrum, and the best qualitative fit
was adopted for the remaining analysis; adopted values are given in
Table \ref{tab.wdfits}.  For most WDs, this adopted fit was the fit
obtained using all Balmer lines.  For LAWDS 5, LAWDS 12, and LAWDS 22,
the adopted fit excluded H$\epsilon$.  For LAWDS 6, both fits were
qualitatively similar, so an intermediate value (also qualitatively
similar) was adopted, and the internal error bars were increased to
include both fit values.  The adopted-fit model Balmer lines are shown
plotted over the observed spectra in Figure \ref{fig.wdfits}.

Masses (\mwd) and cooling ages (\tcool) for each white dwarf were
computed from evolutionary models provided by P.\ Bergeron \& G.\
Fontaine.  These models assume ``thick'' hydrogen layers
($M_H=10^{-4}M_{*}$).  For $\teff\geq 30000$K, the pure carbon cooling
models of \citet{1995LNP...443...41W} are used; for $\teff<30000$K,
the mixed C/O models of \citet{2001PASP..113..409F} are used.  The
evolutionary models also provide synthetic photometry as described in
\citet{2006AJ....132.1221H}.  As pure carbon models are generally
disfavored at these masses, we also calculated \mwd\ and \tcool\ for $\teff\geq
30000$K WDs using the C/O mixed models of \citet{1995LNP...443...41W};
the resulting differences in \mwd\ and \tcool\ were within the stated
error bars.  The derived \mwd\ and \tcool\ for each WD are presented
in Table \ref{tab.wdfits}. Due to the poor fits of LAWDS 5 and LAWDS 6,
their derived masses are of questionable quality.

\subsubsection{Error Determinations}

We consider two primary sources of error in the \teff\ and \logg\
values determined for each WD:

\emph{Internal fitting errors} --- We refer to the random errors in
our determination of \teff\ and \logg\ resulting from the
signal-to-noise of our observations as ``internal fitting errors,'' or
the distribution of spectral parameters we would expect given an
ensemble of observations with the same signal-to-noise of the same WD
using the same instrumental setup.  We determine the magnitude of
these errors via the Monte Carlo technique described in
\citet{2007AJ....133.1490W}.  For LAWDS 5, LAWDS 12 and LAWDS 22,
where the adopted atmospheric parameters used the fits excluding the
H$\epsilon$ line, that line was also excluded in the Monte Carlo error
determination, thereby providing a realistic estimate of the
additional random error introduced by the exclusion.  

\emph{External fitting errors} --- It has been noticed that the
scatter in measured \teff\ and \logg\ derived by different groups for
the same WDs is larger than the internal errors calculated by Balmer
fitting routines
\citep[e.g.,][]{1999ApJ...517..399N,2007AJ....133.1490W}.  Possible
sources for these differences include the different instruments with
which these stars were observed, the different spectral fitting
routines, and the different model atmospheres used.  In
\citet{2007AJ....133.1490W}, we determined that these external errors
appear to be random with a scatter of $\approx 1100$K in \teff\ and a
scatter of $\approx 0.12$ dex in \logg\ based on a comparison of our
observations of field DA WD spectra to fits in the literature. In our
past work \citep[with the exception of][]{Rubin2007}, we have not
propagated the external fitting errors through our analysis.  In order
to facilitate proper comparison of our open cluster WDs with those of
other groups, we propagate these errors through the analysis in this
paper.

We obtain the \emph{total fitting error} in \teff\ and \logg\ for each
WD by adding the internal, and external fitting errors in quadrature.
This total error is then propagated through the analysis to obtain the
stated errors in all subsequently derived parameters (\mwd, \tcool,
and \minit). The internal and total fitting errors are both given in
Table \ref{tab.wdfits}.

\section{The WD population of M35\label{sec.discuss}}
\subsection{Cluster membership\label{sec.discuss.members}}

In the absence of kinematic information, such as radial velocity or
proper motion measurements, the best means of determining cluster
membership of WDs is to apply age and distance criteria.  Any current
cluster member WDs must have \tcool\ shorter than the cluster age and,
unless escaped from the cluster, must lie at the same distance.  Both
\tcool\ and $M_V$ for each WD are determined from the evolutionary
models.  For each WD, \tcool\ is shown (along with other derived mass
quantities) in Table \ref{tab.wdfits}, and $(m-M)_V$ is given along
with other derived photometric quantities in Table \ref{tab.modphot}.
We identify candidate cluster member WDs as those DAs with $\tcool\leq
200$ Myr and that have distance moduli consistent within $2\sigma$ of
the cluster distance modulus.  All of the WDs in this sample meet
these criteria.

\begin{deluxetable*}{lcccccccccccc}
\tablecolumns{13}
\tablewidth{0pt}
\tabletypesize{\scriptsize}
\tablecaption{Model-dependent photometric properties of NGC 2168 white dwarfs\label{tab.modphot}}
\tablehead{ & \multicolumn{6}{c}{Model Photometry} & \multicolumn{6}{c}{Derived Distance and Reddening}\\
\cline{3-7}\cline{9-13}
\colhead{WD} & \colhead{$M_V$} & \colhead{$\sigma_{M_V}$} & \colhead{$_{(\bv)_0}$} & \colhead{$\sigma_{(\bv)_0}$} & 
  \colhead{$_{(\uv)_0}$} & \colhead{$\sigma_{(\uv)_0}$} & \colhead{$_{(m-M)_V}$} & \colhead{$\sigma_{(m-M)_V}$} & 
  \colhead{$_{\ebv}$} & \colhead{$\sigma_{\ebv}$} & \colhead{$_{E(\uv)}$} & \colhead{$\sigma_{E(\uv)}$}}
\startdata
  1 & 10.300 &  0.283 & -0.221 &  0.011 & -1.420 &  0.016 & 10.689 &  0.284 &  0.186 &  0.030 &  0.290 &  0.043 \\
  2 & 10.736 &  0.254 & -0.222 &  0.010 & -1.433 &  0.014 & 10.833 &  0.256 &  0.283 &  0.045 &  0.397 &  0.062 \\
  5 &  9.759 &  0.247 & -0.291 &  0.003 & -1.563 &  0.004 & 10.306 &  0.248 &  0.163 &  0.024 &  0.262 &  0.035 \\
  6 &  9.592 &  0.319 & -0.295 &  0.003 & -1.570 &  0.004 & 10.271 &  0.319 &  0.167 &  0.023 &  0.275 &  0.034 \\
 11 & 11.090 &  0.270 & -0.051 &  0.025 & -1.006 &  0.045 & 10.108 &  0.271 &  0.161 &  0.045 &  0.436 &  0.069 \\
 12 & 10.382 &  0.300 & -0.228 &  0.009 & -1.438 &  0.013 & 10.921 &  0.301 &  0.273 &  0.039 &  0.474 &  0.055 \\
 14 & 11.041 &  0.264 & -0.169 &  0.016 & -1.324 &  0.025 & 10.660 &  0.265 &  0.228 &  0.041 &  0.450 &  0.058 \\
 15 & 10.622 &  0.270 & -0.179 &  0.015 & -1.330 &  0.025 & 10.163 &  0.271 &  0.140 &  0.035 &  0.272 &  0.052 \\
 27 & 11.008 &  0.265 & -0.182 &  0.014 & -1.353 &  0.021 & 10.390 &  0.266 &  0.272 &  0.041 &  0.434 &  0.059 \\
 29 & 10.393 &  0.246 & -0.212 &  0.011 & -1.401 &  0.017 & 10.326 &  0.247 &  0.137 &  0.029 &  0.239 &  0.042 \\
 30 & 10.957 &  0.428 & -0.185 &  0.016 & -1.358 &  0.024 & 10.218 &  0.428 &  0.180 &  0.033 &  0.426 &  0.048 \\
\enddata
\end{deluxetable*}

In \citet{2004ApJ...615L..49W}, we suggested that LAWDS 15 could also
be an escaped cluster member. After the re-analysis
presented in this paper, the WD distance modulus is consistent with
the cluster distance modulus, so the speculation on its potential
escape from the cluster was premature.

There are some factors that lead us to question the membership of DA
LAWDS 11. Its cooling age is $\sim 40$ Myr older than the
second-oldest WD, LAWDS 14, though this difference is within the
stated errors.  In addition, LAWDS 11 is the second-least massive WD
in the cluster, despite having the highest progenitor mass.  This goes
against the preconceived notion that higher-mass stars produce
higher-mass WDs.  There are theoretical means to explain this low
mass, such as binary evolution or enhanced mass loss during the
post-main sequence evolution of the star.

It is also possible that LAWDS 11 is a field WD unrelated to the star
cluster that lies within our selection volume.  Using a calculation
similar to that described in \citet{Rubin2007}, we estimate $\approx
3.2$ field WDs meeting our cluster-member criteria should be in our imaged
area. As $\sim 22\%$ of field WDs in a volume-limited sample are more
massive than 0.8\msun \citep{2005ApJS..156...47L}, it would not be
improbable to find a massive field WD in our sample.  Measurements of
this star's proper motion are needed to determine if it is indeed a
member of the cluster; as LAWDS 11 meets all of our current membership
criteria, we include it in our analysis below.

\subsubsection{Sample Completeness}

The fourteen WDs observed in this cluster may not be the complete WD
population of M35.  Our imaging is only of the central $\approx 18$
arcmin radius of the cluster; the tidal radius of M35 is $\gtrsim 33$
arcmin \citep{1989ApJ...339..195L}.  Therefore, it is fully possible
that cluster WDs may be found outside our imaged area.  However,
almost all main-sequence proper motion members more massive than
1.2\msun\ are located within a 20 arcmin radius of the cluster center
\citep{1986ApJ...310..613M,2001ApJ...546.1006B}, and the progenitor
stars of M35 WDs were significantly more massive than this, suggesting
that few cluster member WDs may be outside of our images.

We may also be missing cluster WDs within our imaged area.  Because
precise WD parameter measurements rely on the higher-order Balmer
lines and because an atmospheric dispersion corrector was not
available when these observations were made, we decided to make
longslit observations of individual WDs at parallactic angle rather
than multi-slit observations at less favorable angles. Therefore, most
of the WD candidates lack spectroscopic observations.

Many of the unobserved WD candidates lie along and above
the 0.6\msun\  WD cooling curve; all of the observed DAs are more
massive than this and, with the exception of LAWDS 22 (see
\S\ref{sec.spec.obs}), lie along more massive WD cooling curves. We
cannot rule out that the unobserved candidates may be cluster WDs in
double degenerate or WD+M binaries, or low-mass cluster WDs formed via
some binary process. While interesting in their own right, any of
these scenarios would warrant their exclusion from the initial-final
mass relation and \mcrit\ analyses below, which are assumed to be
valid only for single-star evolution.  

The other unobserved WD candidates, if cluster members, would have
cooling ages $\geq 200$ Myr, the oldest likely age for the cluster.
Therefore, they are almost certainly not white dwarfs related to the
star cluster. 

Most of the WD candidates lacking spectroscopic observations are
likely field WDs. As mentioned in the previous section, we estimate
that, on average, 3.2 field WDs meeting our cluster membership
selection criteria should be found within our field of view. Our
photometric selection criteria are much more generous, consistent with
WDs as cool as $\sim 7000$K at the cluster distance and reddening;
such WDs would be significantly older than the star cluster, and
therefore not cluster members.  Solely using our photometric selection
criteria and integrating along the line of sight to the cluster, we
estimate $\sim 35$ field WDs to be in our field of view. Given our 41
total WD candidates, 10 to 14 of which are cluster members, we expect
that most or all of these unobserved WD candidates are field WDs.

In the \bv\ vs. $V$ CMD (Fig.~\ref{fig.id_cmd}), a tight, nearly
continuous sequence of member WDs is observed.  There are no remaining
WD candidates along the 1\msun\ cooling track younger than 200 Myr,
suggesting that no older, massive WDs are present in the
cluster. However, we cannot rule out that cooler WDs were formed and
lost from the cluster due to dynamical processes.  Photometric
incompleteness is also a concern; artificial star tests (significantly
complicated by proper treatment of the mosaicked CCDs) are being
performed and will be presented in a future paper.

In short, spectroscopic observations of the unobserved WD candidates
are needed to ascertain their true nature, but they are quite unlikely
to impact the conclusions of this paper.

\subsection{The initial-final mass relation\label{sec.discuss.ifmr}}

The progenitor star masses for cluster WDs can be determined by
subtracting \tcool\ from the cluster age.  The result represents the
lifetime of the progenitor star from the zero-age main sequence
through the planetary nebula stage.  We then use the [Fe/H]$=-0.2$
evolutionary models of \citet{2007A&A...469..239M} to determine the
mass of star with this lifetime.  These model lifetimes only extend
from the zero-age main sequence through the start of the
thermally-pulsing AGB phase.  The amount of time from the first
thermal pulse through the thermally-pulsing AGB and planetary nebula
phases to the start of the WD cooling track is negligible ($\sim 10^5$
yr) compared with the total stellar lifetimes ($\sim 10^8$ yr) and the
errors on the cluster age ($\sim 10^7$ yr).

\begin{figure}
\plotone{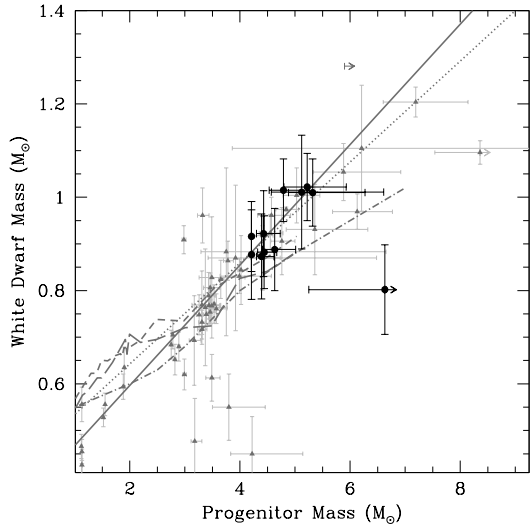}
\caption{The empirical initial-final mass relation.  Points for M35
  (filled circles) are shown along with cluster WDs from the
  literature (triangles with gray error bars and lower limits). Errors
  are 68\% confidence limits and include only the propagated total
  fitting errors, not uncertainties in cluster ages. Curves are linear
  fits to the IFMR as determined by this work (solid line) and
  \citet[dotted line]{2005MNRAS.361.1131F}, the semi-empirical IFMR of
  \citet[dot-dashed curve]{2000A&A...363..647W}, and the core mass at
  the end of thermal pulsing for $Z=0.008$ and $Z=0.019$ AGB stars
  \citep[short- and long-dashed lines,
  respectively;][]{2007A&A...469..239M}. \label{fig.ifmr}}
\end{figure}

Table \ref{tab.wdfits} gives the progenitor mass \minit\ for each
cluster WD.  This table excludes parameters for LAWDS 22 due to
its likely magnetic nature and resulting poor spectral fits.
The table lists the observational errors in \minit\ resulting from the
propagation of the total fitting errors through the calculations.
Also given are the systematic errors, the changes in \minit\ due to
the 25 Myr uncertainty in the cluster age. These systematic errors
would be in the same sense for each cluster WD and so should not be
added in to the random error (see \S\ref{sec.discuss.errors}).  The
given errors in \minit\ are strongly asymmetric and represent the 68\%
confidence level.  The initial-final mass relation is plotted in
Figure \ref{fig.ifmr}.  Error bars are the 68\% confidence levels
derived from the total fitting errors.  Despite the uncertainty in the
spectral fits for LAWDS 5 and LAWDS 6, their \minit\ values and
uncertainties are fairly robust; as their \tcool\  are very short,
even an error in \teff\  of 10000 K  translates to an error in \minit\  of
only 0.02\msun.

Also included in Figure \ref{fig.ifmr} are cluster WDs from the
literature.  For consistency, we re-determined \minit\ and \mwd\ from
the published \teff\ and \logg\ using our adopted WD and stellar
evolutionary models. These clusters, the adopted parameters from each
cluster, and references for the parameters and WD observations are
given in Table \ref{tab.cldata}.  Points from the literature with
small error bars typically do not include the external fitting errors
that we have considered in our data.

\begin{deluxetable}{lccc}
\tablecolumns{4}
\tablewidth{0pt}
\tabletypesize{\footnotesize}
\tablecaption{Adopted cluster parameters for previously-published
  white dwarfs\label{tab.cldata}}
\tablehead{\colhead{Cluster} & \colhead{Age (Myr)} & \colhead{[Fe/H]} &
  \colhead{References}} 
\startdata
Pleiades  & $125\pm 25$ & $ 0.00$ & 1,2,3,4,5 \\
NGC 2516  & $158\pm 20$ & $-0.10$ & 6,7,8 \\
M34       & $225\pm 25$ & $ 0.07$ & 9,10,11 \\
Sirius A/B& $238\pm 13$ & $ 0.00$ & 12,13 \\
NGC 2287  & $243\pm 40$ & $ 0.00$ & 14 \\
NGC 3532  & $300^{+25}_{-25}$ & $ 0.00$ & 6,14 \\
NGC 2099  & $490\pm 70$ & $ 0.05$ & 14,15,16 \\
NGC 6633  & $562^{+69}_{-61}$ & $-0.10$ & 17,18\\
Hyades    & $625\pm 50$ & $ 0.13$ & 3,19,20 \\
Praesepe  & $625\pm 50$ & $ 0.13$ & 3,21,22\\
NGC 7789  & $1400\pm  140$ & $-0.10$ & 23 \\
NGC 6819  & $2500\pm  250$ & $-0.02$ & 23 \\
NGC 6791  & $8500\pm 1000$ & $ 0.35$ & 24,25 \\
\enddata
\tablerefs{(1) \citealt{1989A&A...213L...1M}, (2)
  \citealt{1998ApJ...499L.199S}, (3) \citealt{2001ApJ...563..987C},
  (4) \citealt{2005MNRAS.361.1131F}, (5)
  \citealt{2006MNRAS.373L..45D}, (6) \citealt{1993A&AS...98..477M},
  (7) \citealt{1996A&A...313..810K}, (8)
  \citealt{2002AJ....123..290S}, (9) \citealt{1996AJ....111.1193J},
  (10) \citealt{2003AJ....125.2085S}, (11) \citealt{Rubin2007}, (12)
  \citealt{2005ApJ...630L..69L}, (13) \citealt{2005MNRAS.362.1134B},
  (14) \citealt{Dobbie2008}, (15) \citealt{2005AJ....130.1916M},
  (16) \citealt{2008ApJ...675.1233H}, (17)
  \citealt{2002MNRAS.336.1109J}, (18) \citealt{2007AJ....133.1490W},
  (19) \citealt{1990ApJ...351..467B}, (20)
  \citealt{1998A&A...331...81P}, (21) \citealt{2004MNRAS.355L..39D},
  (22) \citealt{2006MNRAS.369..383D}, (23)
  \citealt{2007arXiv0706.3894K}, (24) \citealt{2006ApJ...646..499O},
  (25) \citealt{2007ApJ...671..748K} }
\end{deluxetable}

As we are seeking the cleanest-possible IFMR indicative of the results
of single-star evolution, a few WDs from the literature are excluded
from our analysis.  WD 0437+138 is of spectral type DBA; the cooling
models we employ are appropriate only for DA WDs.  WD 0837+199 is
strongly magnetic, which could potentially indicate a more complicated
evolutionary history \citep[e.g.,][]{2008MNRAS.387..897T}.  The two
binary WD candidates from \citet{2007AJ....133.1490W}, NGC 6633:LAWDS
4 and NGC 6633:LAWDS 7, either have uncertain parameters (if binaries,
the observed spectra are a blend of the two components) or, if lone
WDs, are not cluster members.  Proper motion studies have identified
the following WDs as field stars: WD 0837+218 \citep{Casewell2008},
NGC 1039:LAWDS 9, NGC 1039:LAWDS 20, NGC 1039:LAWDS S1, and NGC
7063:LAWDS 1 \citep{Dobbie2008}.

From the figure, we see that the M35 WDs fall along the existing
empirical IFMR, with higher cluster WD masses are seen to correspond
to higher progenitor star masses, as is generally expected from
stellar evolution.  In fact, the M35 WDs follow the linear relation of
\citet{2005MNRAS.361.1131F}, perhaps with a slightly steeper slope.
Also included in the figure are the semi-empirical IFMR of
\citet{2000A&A...363..647W} and a theoretical prediction for the IFMR
from \citet{2007A&A...469..239M} for $Z=0.008$ and $Z=0.019$ stellar
evolutionary models.

A recent paper by \citet{2008arXiv0807.3567S} highlights that the IFMR
determined via our method is not completely self-consistent.  The
cluster age was adopted from literature using isochrone-fitting
methods employing different stellar evolutionary models; the WD
evolutionary models do not use the chemical profiles output by the
same stellar models used to determine the cluster age; and the
atmospheric models used to fit the spectra are not identical to those
used in the WD evolutionary models.  However, as pointed out by
\citet{2008arXiv0807.3567S}, the overwhelming source of systematic
error in IFMR calculations is from the uncertainty in the cluster age
(see \S\ref{sec.discuss.errors.age}).  The systematic errors due to
the self-inconsistency described above is therefore assumed to be negligible.

\subsubsection{The shape of the IFMR}

The empirical IFMR can be fit marginally well by a linear function.
We determine the slope and zero-point of this line using least-squares
fitting with error bars in two dimensions \citep[e.g.,][]{Recipes}. We
include the 61 WDs in Figure \ref{fig.ifmr} with masses $M_{\rm
WD}\geq 0.5\msun$; lower masses, which may indicate He cores and a
different evolutionary history, were omitted.
LAWDS 11 was also omitted from the fit due to its uncertain
membership; if it is a cluster member and is the result of binary
evolution, this exclusion would also be justified.  Using the
published random error bars for each WD's parameters, we obtain
\begin{equation}
M_{\rm final} = 0.339\pm0.015 + (0.129\pm0.004) \minit\,;
\end{equation}
this fit is plotted in Figure \ref{fig.ifmr}.  The reduced $\chi ^2$
value for this fit is 2.47.  We note that, if LAWDS 5 and LAWDS 6 are
also excluded from the fit due to their poor spectral fits, the linear
fit is unchanged within the formal error bars.

Our linear fit is formally inconsistent with the slopes of the fits
given in \citet{2005MNRAS.361.1131F}, \citet{2007arXiv0706.3894K} and
\citet{2008MNRAS.387.1693C}; the M35 WDs at the high-mass end of the
relation prefer a steeper slope.
However, the goodness-of-fit statistic for all of these linear fits
(including the new fits presented here) are very low.  This could
indicate that the error bars on the WD parameters are understated,
that there is an additional scatter not due to measurement error (such
as intrinsic scatter in the relation), and/or that a linear function
is not an appropriate representation of the data
\citep[e.g.,][]{Recipes}.  In any of these cases, the formal error
bars on the slope and zero point of the linear fit quoted above are
likely understated. Qualitatively, all of these linear fits appear to
be a decent approximation to the data.

We cannot exclude the possibility that the empirical IFMR may not be
strictly linear; emergent hints of a non-linear IFMR have been discussed
recently in the literature
\citep[e.g.][]{2007arXiv0706.3894K,2008arXiv0807.3567S,2008MNRAS.387.1693C,Dobbie2008}. The
formal significance of these non-linear fits is limited by the larger
error bars in massive WDs (driven by cluster age uncertainties and
the steep change in the relationship between stellar mass and
lifetimes at these younger ages) and the limited amount of data at the
low \minit\ end of the relation
\citep[e.g.,][]{2008MNRAS.387.1693C,2008arXiv0807.3567S}.

\subsection{Systematic errors in the IFMR\label{sec.discuss.errors}}

\subsubsection{The age uncertainty in M35\label{sec.discuss.errors.age}}

Although several systematic effects may affect the IFMR, by far the
largest systematic in M35 is the uncertainty in the cluster age.
Errors in the cluster age affect the IFMR in two distinct ways.
First, an increase (decrease) in the assumed cluster age will result
in \emph{all} WDs from a given cluster having lower (higher) derived
initial masses.  Second, a change in the cluster age results in a
larger shift for cluster WDs with higher progenitor masses, as the
change in age represents a larger fraction of the main sequence
lifetime for these stars.

\begin{figure*}
\mbox{
\includegraphics[width=0.3\textwidth]{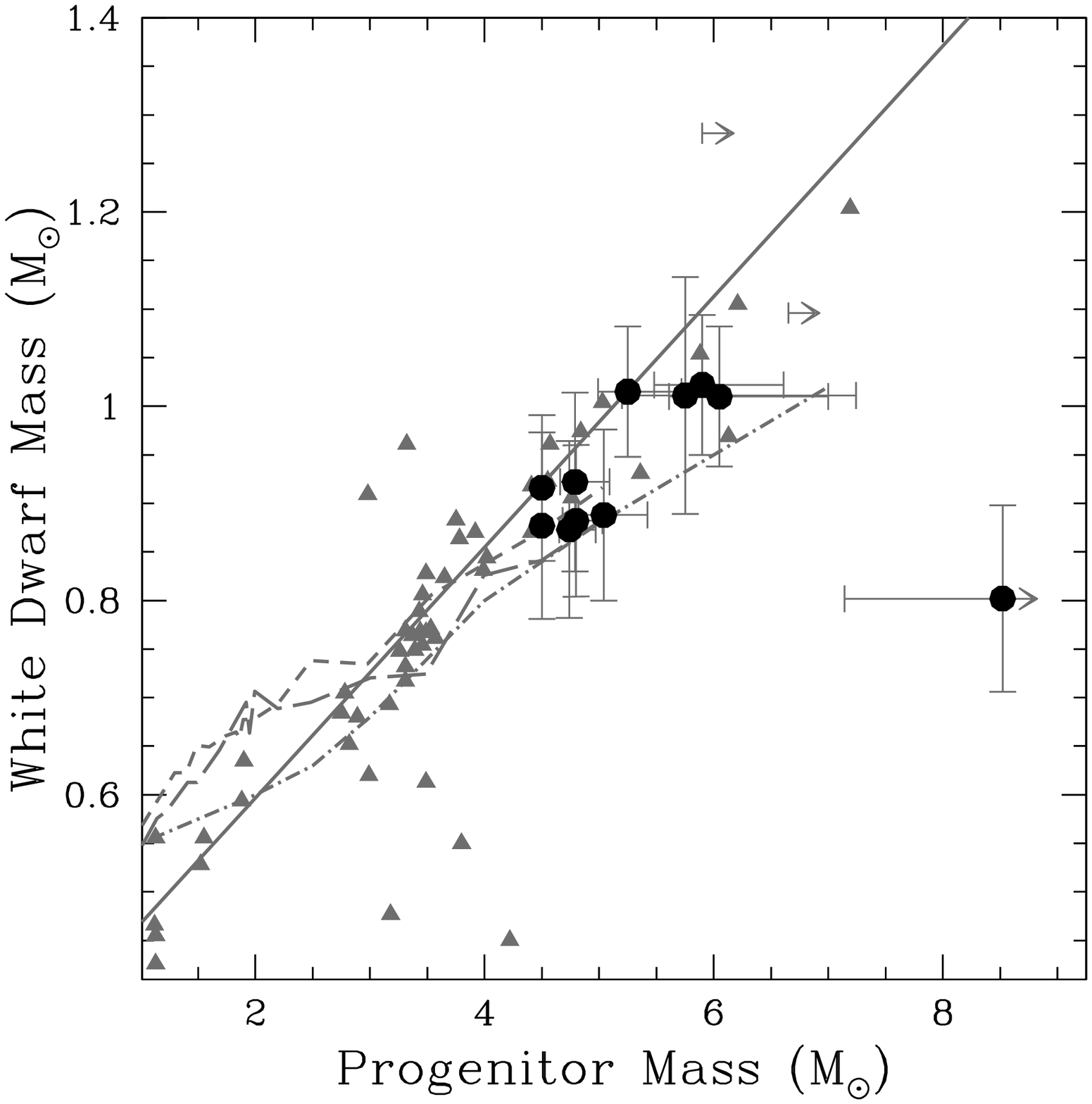}
\includegraphics[width=0.3\textwidth]{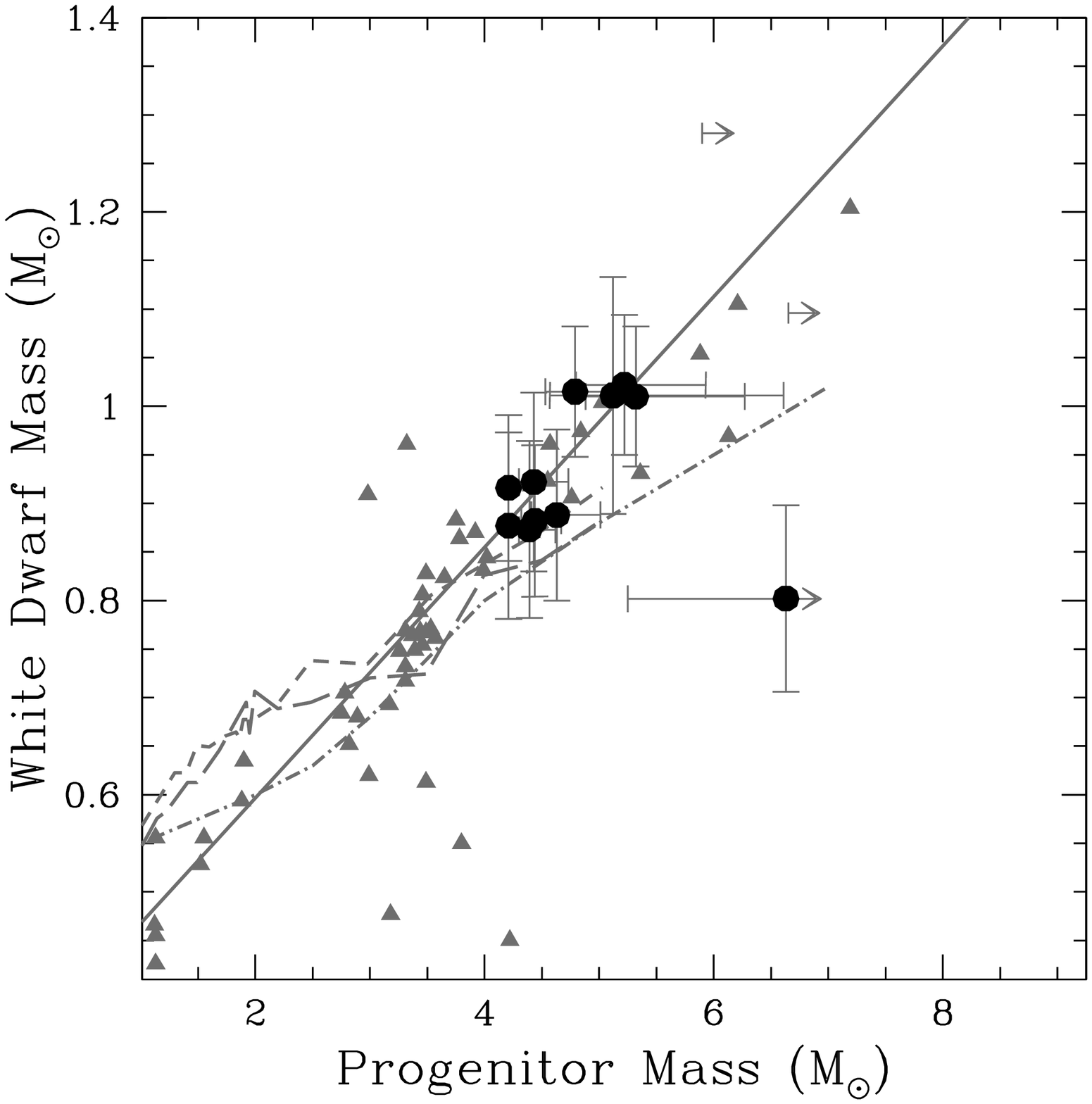}
\includegraphics[width=0.3\textwidth]{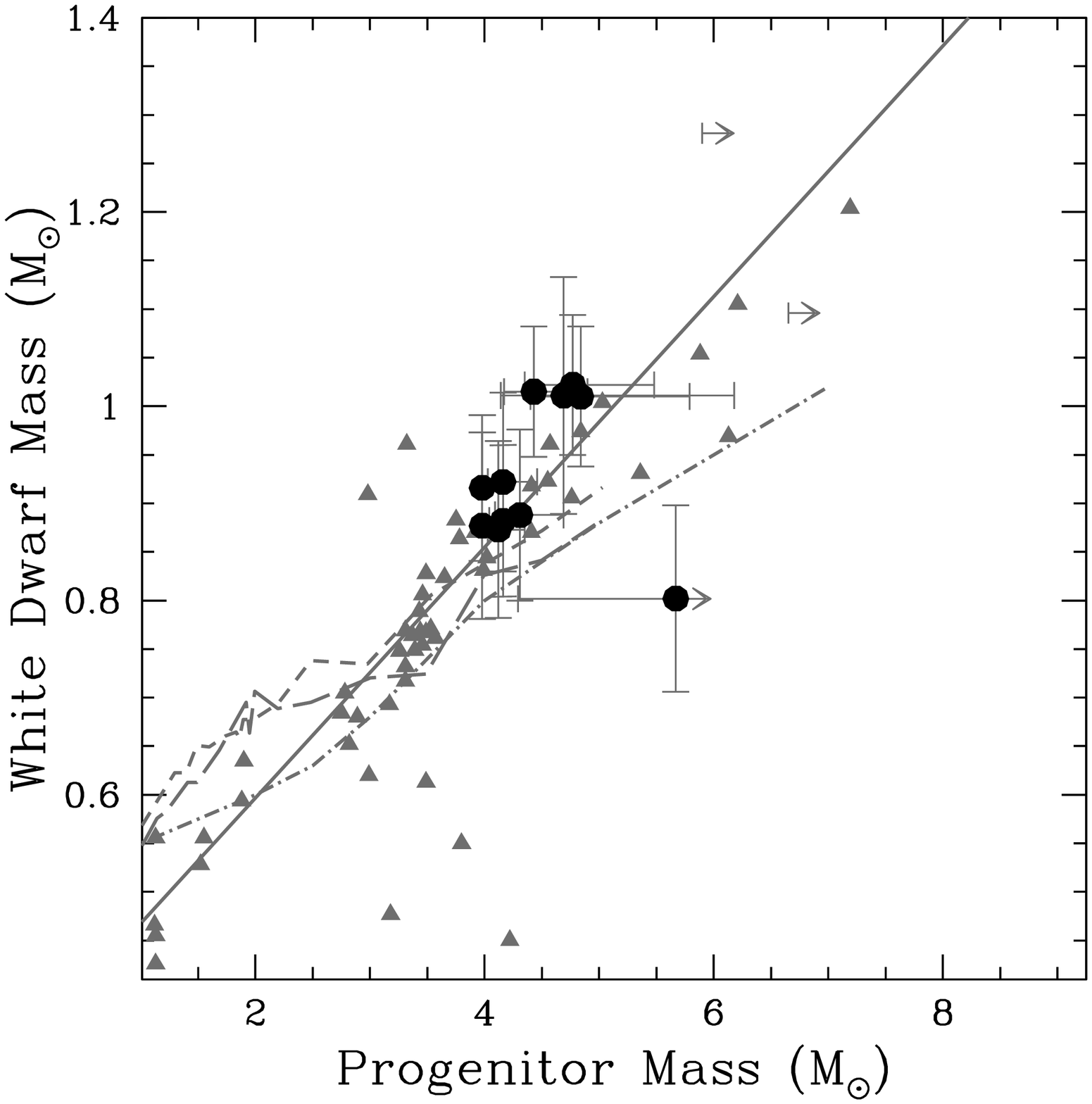}
}
\caption{The effects of assumed cluster age on the empirical
  initial-final mass relation.  The M35 points shift systematically to
  lower $M_{\rm init}$ and their slope steepens as the assumed cluster
  age is increased from (left) 150 Myr to (center) 175 Myr and (right) 200 Myr.
  Symbols are as in Figure \ref{fig.ifmr}. \label{fig.ifmr_age}}
\end{figure*}

Both of these effects can be illustrated in the M35 empirical IFMR.
Figure \ref{fig.ifmr_age} shows the empirical IFMR for three different
assumed ages of M35: 150 Myr, 175 Myr, and 200 Myr.  The figure
illustrates that, for older assumed ages, the cluster points shift
systematically to lower \minit.  The slope of the cluster IFMR also
steepens significantly with increasing assumed cluster ages.

The conclusions drawn from comparison of the M35 IFMR to that of other
clusters will differ depending on the assumed cluster age.  If the
cluster age is 150 Myr, the cluster IFMR agrees extraordinarily well
with the empirical IFMR derived from other clusters.  Yet if the
cluster age is 200 Myr, one would conclude that the cluster WDs are
systematically more massive than those in other young clusters,
perhaps indicative of metallicity-dependent mass loss rates
\citep[e.g.,][]{2007A&A...469..239M,2007ApJ...671..748K}.

These same systematics affect attempts to calculate the intrinsic
scatter in the IFMR.  For a younger M35, the scatter at a given
\minit\ would be explained by the intrinsic observational errors.  Yet
for an older cluster, the intrinsic scatter would be significant.

This systematic is most severe in young clusters; for old
open clusters, the uncertainty in the cluster age is small relative to
the progenitor star lifetimes. For example, the 1 Gyr ($\sim 15\%$)
uncertainty in the age of NGC 6791 results in just a 0.05\msun\ change
in \minit\ for its WDs, compared with the $\gtrsim 1\msun$ uncertainty
in \minit\ for the M35 WDs for the cluster age uncertainty of just 25
Myr (also $\sim 15\%$).  


\subsection{Limits on \mcrit\label{sec.discuss.mcrit}}

Observational lower limits on the maximum mass of WD progenitors,
\mcrit, can be determined from IFMR.  To zeroth order, this can be
accomplished by simply identifying the WD with the highest \minit\ in
the empirical IFMR.  However, this approach is complicated by numerous
uncertainties.  First, the sizes of the error bars on individual WDs
are large enough that it is not possible to identify the single WD
with the highest \minit.  Second, the errors on \minit\ of individual
WDs are asymmetric and non-analytic, complicating any effort to simply
calculate limits on \mcrit.  Third, the WDs with the highest \minit\ 
come from numerous star clusters, each of which will have its own
unique systematic errors in \minit\ due to the cluster age
uncertainties.  We therefore elect to use a Monte Carlo simulation to
determine the lower observational limits on \mcrit.

The simulation creates 25000 realizations of the observed WD
population in the open clusters M35, the Pleiades \citep[including GD
50 and PG 0136+251, ][]{2006MNRAS.373L..45D}, NGC 2516, NGC 3532,
NGC 2287, and Sirius B; WDs from older open clusters are excluded
because they either have such low \minit\ or such large error bars
that they do not contribute to the limits on \mcrit.  For each
realization, each open cluster is a assigned an age randomly drawn
from the values quoted in Table \ref{tab.cldata}; as the published
errors in cluster ages tend to indicate acceptable ranges rather than
a normal distribution about the best value, cluster ages are
drawn randomly from a uniform distribution.  Additionally, in each
realization a \teff\ and \logg\ are assigned to each known WD, with
values drawn from a normal distribution with a mean of the observed
\teff\ and \logg\ for the WD, and a standard deviation equal to the
stated random errors in each quantity.The assigned \teff\ and \logg\
are then used to determine \minit\ for each WD; if the cooling age of
the simulated WD is older than the cluster age, that WD is ignored for
that particular realization.  For each realization, the highest
overall \minit\ is identified as the lower limit on \mcrit\ for that
realization.

We perform six runs on this Monte Carlo simulation with varying
parameters.  Three different WD populations are tested: 
\begin{itemize}
\item Only the WDs in NGC 2168; this limits the number of systematics
involved in the simulation, but provides the loosest limits on
\mcrit,
\item All DA WDs in the five open
clusters mentioned in the previous paragraph as well as Sirius B; this
data set represents all available useful cluster data, and
\item The previous sample minus three WDs of uncertain cluster membership:
NGC 2168 LAWDS 11 (see \S
\ref{sec.discuss.members}), and GD 50 and PG
0136+251, as their common space motion with the Pleiades does not
prove they were born simultaneously with the cluster. 
\end{itemize}
Each of these three WD populations was paired with two variations on
WD core composition.  In the first instance, all WDs are assumed to
have carbon-oxygen cores; in the second, any WDs with $\mwd\geq
1.05\msun$ are assumed to have oxygen-neon cores (see
\S\ref{sec.discuss.massive}).

\begin{deluxetable*}{lcccccccc}
\tablecolumns{9}
\tablewidth{0pt}
\tablecaption{Lower Limits on \mcrit\label{tab.mcrit}}
\tablehead{\colhead{WD Sample} & \multicolumn{4}{c}{Carbon/Oxygen} 
& \multicolumn{4}{c}{Oxygen/Neon} \\
  & \colhead{50\%} & \colhead{90\%} & \colhead{95\%} & \colhead{99\%}
  & \colhead{50\%} & \colhead{90\%} & \colhead{95\%} & \colhead{99\%} \\
  & & \colhead{(\msun)} & \colhead{(\msun)} & \colhead{(\msun)} & \colhead{(\msun)}
  & \colhead{(\msun)} & \colhead{(\msun)} & \colhead{(\msun)}}
\startdata
M35 alone    & 6.57 & 5.40 & 5.19 & 4.89 & 6.48 & 5.34 & 5.14 & 4.86 \\
All Clusters & 8.86 & 7.41 & 7.08 & 6.51 & 8.39 & 6.95 & 6.70 & 6.34 \\
All Clusters, ``Cleaned'' & 7.97 & 6.58 & 6.25 & 5.77 & 7.80 & 6.54 & 6.30 & 5.91\\
\enddata
\tablecomments{Masses are lower limits on \mcrit\ from Monte Carlo
  simulations for the given
  confidence interval and given core
  composition for $M_{\rm WD}\geq 1.05\msun$ (lower mass WDs are
  assumed to have C/O cores). The ``cleaned'' sample excludes LAWDS
  11, GD 50, and PG $0136+251$}
\end{deluxetable*}

Results of these calculations are presented in Table \ref{tab.mcrit}.
Based on the WDs in M35 alone, the 95\% confidence lower limit on
\mcrit\ is $5.1\msun-5.2\msun$, depending on assumed core composition.
By combining all available cluster WD data, the 95\% confidence lower
limit on \mcrit\ increases to $6.3\msun-7.1\msun$, depending on the
membership of LAWDS 11, GD 50, and PG 0136+251 and on the assumed core
composition.

We also note that the combination of \emph{all} cluster data is
required to get the tightest limits on \mcrit.  For example, the
Pleiad WD, LB 1497, has the highest plotted \minit\ in
Fig.~\ref{fig.ifmr}. In our Monte Carlo calculations excluding LAWDS
11, GD 50, and PG 0136+251, LB 1497 only provides the highest \minit\
in 38\% of the realizations.  Messier 35 WDs provide the highest
\minit\ in 42\% of the realizations, and NGC 2516 in 19\% of the
realizations.  Excluding any one of these three clusters from the
Monte Carlo calculation lowers the limit on \mcrit\ by $\sim 0.5\msun
- 1\msun$. 

Combining these lower limits on \mcrit\ with emerging upper limits
from supernova progenitor searches \citep[$\mcrit\lesssim 9.5\msun$,
e.g.,][]{2006MNRAS.369.1303H,Smartt2008}, it may be reasonable to
assert that $6\msun \leq \mcrit \leq 9.5\msun$ with 95\% confidence
for stars near solar-metallicity.  However, a more careful analysis of
both the WD and supernova data is needed before these constraints can
be claimed with confidence.

\subsubsection{Systematic errors in the limits on \mcrit}

The primary sources of systematic error in these simulations are from
the input star cluster ages and the input stellar evolutionary models.
The input ages are all determined from evolutionary models including
moderate convective overshoot, and so are reasonably consistently
determined despite coming from a multitude of published sources and
methods.  However, there is a well-known degeneracy between an
observed cluster's age, metallicity, and distance; a significant
change in the age of any of the clusters used in this simulation would
change the results.  And, as stellar models continue to evolve, the
derived ages for the entire cluster sample may change systematically.

These same evolutionary models are used to convert the WD progenitor
lifetime into \minit.  Therefore, any significant changes in the
output lifetimes for intermediate-mass stars would also significantly
impact these results.  However, moderate levels of convective
overshoot have become generally accepted in the community and appear
to be borne out by observations
\citep[e.g.,][]{2007A&A...475.1019C,2008AJ....135.1757S}.

How large a difference could future changes to evolutionary models
make?  According to the Padova isochrones, a solar-metallicity model
including moderate convective overshoot with a lifetime of 100 Myr has
a mass of 5.42\msun, and a mass of 5.18\msun\ with no convective
overshoot.  Therefore, it seems unlikely that our derived limits will
change by $\lesssim 0.3\msun$ due to a choice of evolutionary models
alone.  Larger changes are possible, though, depending on how newer
evolutionary models affect the input star cluster ages.

Another possible source of error is due to our combination of data
from star clusters of differing metallicity; namely, NGC 2168 is
significantly subsolar metallicity, while the remaining clusters and
Sirius are close to solar metallicity.  Numerous models of super-AGB
stars exhibit a dependence of \mcrit\ on metallicity \citep[e.g.,][and
references therein]{2007A&A...476..893S}.  Though these models differ
significantly on the precise value of \mcrit, they predict that
\mcrit\ for stars with the metallicity of NGC 2168 should be
systematically $\sim 0.2\msun$ to 0.3\msun\ lower than for
solar-metallicity stars.

We note that incompleteness in the WD sample cannot reduce our lower
limits on \mcrit; additional WDs can only raise the lower limit.

\subsection{The cluster white dwarf distance  modulus\label{sec.discuss.dmod}} 

The distance modulus and reddening we have adopted for M35
$[(m-M)_V=10.3\pm 0.1$, $\ebv=0.22\pm 0.03]$ are based solely on
previously-published main-sequence fitting derivations.  As the
spectral fits to each WD, when combined with the WD photometry,
provide a measure on the distance modulus and reddening for each WD,
we can use the ensemble averages of these individual measurements as
an independent determination of the cluster distance and reddening.
Again, due to the uncertainty in its spectral parameters, we do not
include LAWDS 22 in these calculations.

The weighted means of the individual distance moduli give a mean WD
distance modulus of $(m-M)_V=10.45\pm 0.08$, with a dispersion of
$\sigma_{(m-M)}=0.28$ mag. The mean color excesses are $\ebv=0.185\pm
0.010$, $\sigma_{E(B-V)}=0.055$ and $E(U\!-\!V)=0.329\pm 0.014$,
$\sigma_{E(U-V)}=0.091$.  The mean cluster WD distance modulus and
\ebv\ are therefore fully consistent with the values determined from
main-sequence fitting; the ratio $E(U\!-\!V)/\ebv=1.78\pm 0.12$ is
consistent with the Milky Way reddening law of
\citet{1985ApJ...288..618R} for $R_V=3.1$.

The measured scatter in the measured color excesses is larger than the
uncertainties in the individual measurements (see Table
\ref{tab.modphot}). In addition, there is a significant correlation
between the individual WD distance moduli and the \ebv\  color excess.
These two effects could be explained by differential interstellar
extinction across the cluster. If we ``correct'' each distance modulus
to the mean \ebv\  color excess, the dispersion in the distance moduli
drops to $\sigma_{(m-M)}=0.19$ mag.

However, there is no obvious spatial correlation between WDs with
similar measured \ebv.  In addition, errors in the measured \teff\  will
mimic extinction effects.  If the measured \teff\  is higher than the
actual value, then the derived WD absolute magnitude will be too bright,
resulting in too large of a measured distance modulus.  At the same
time, the derived model color will be bluer than the actual WD,
resulting in an artificially large measured \ebv. The magnitude of
this effect is essentially identical to the stated errors in the \ebv\ 
and $E(U\!-\!V)$ measurements in Table \ref{tab.modphot}.  Therefore, the
errors in \teff\  can explain most, but not all, of the observed scatter
in the \ebv\  and $E(U\!-\!V)$ values.

\subsection{Notes on interesting objects}
\subsubsection{Potential oxygen-neon core WDs\label{sec.discuss.massive}}

Four M35 WDs, LAWDS 2, LAWDS 14, LAWDS 27, and LAWDS 30, have masses
within $1\sigma$ of 1.05\msun, the lower mass of ONe core WDs that may
be produced by super-AGB stars \citep{2007A&A...465..249A}.  For this
reason, we have also calculated \mwd\ and \minit\ for these stars
using the ONe WD evolutionary models of \cite{2007A&A...465..249A}.
The ONe models reduce \minit\ by $\approx 0.2\msun$ for all four WDs.
The decrease in \minit\ is due primarily to the lower heat capacity of
ONe cores, which allow the ONe-core WDs to cool more rapidly than
C/O-core WDs of the same mass.

From our data, we have no means of knowing if these two WDs have 
carbon-oxygen or oxygen-neon cores.  However, stellar evolutionary
models of the $\sim 5\msun$ progenitors of these WDs suggest that they
should have C/O cores. 

\subsubsection{The DB white dwarf LAWDS 4 \label{sec.discuss.db}}

Claims have been made that a lack of DB WDs exists in younger open
clusters \citep{2005ApJ...618L.129K,2007arXiv0706.3894K}. Several
non-DA WDs are known to exist in the field of younger open clusters,
including LP 475-252 (spectral type DBA in the
Hyades), NGC 2168: LAWDS 28 \citep[the hot DQ in this cluster;
][]{2006ApJ...643L.127W}, NGC 1039:LAWDS 26, and NGC
6633:LAWDS 16 \citep{2007AJ....133.1490W,Rubin2007}, though only LP
475$-$252 has been confirmed as a cluster member via proper motion.
We have identified one DB WD in the field of M35, LAWDS 4.  If this is
a cluster member, it would make M35 the youngest open star cluster
with a DB WD member (noting again that M35 contains the
non-DA LAWDS 28).

Is LAWDS 4 a cluster member?  At present, the low signal-to-noise of
the spectrum precludes a robust spectral fit.  However, we can
estimate its \teff.  The weakness of the helium absorption lines, the
spectral slope, and the colors are all consistent with a fairly cool
($\teff\approx 15000-17000$K) DB WD.  Based on cooling models provided
by P.\ Bergeron, this DB could only be this cool and still be younger
than M35 if it has a relatively low mass ($M\lesssim 0.6\msun$).

If we assume a WD mass of 0.4\msun\ to 0.6\msun\ and $\teff= 17000$K,
the apparent distance modulus of LAWDS 4 is between 10.24 and 10.77,
consistent with the cluster distance modulus.  In other words, this DB
is photometrically consistent with cluster membership, under the
assumption that its mass is $M\lesssim 0.6\msun$.

If LAWDS 4 is a cluster member, its progenitor mass would have to be
high ($\gtrsim 5\msun$), yet the WD mass ($\lesssim 0.6\msun$) lies
well below the empirical initial-final mass relation, possible under a
binary formation scenario for the WD.  It is also possible, and
perhaps probable, that this star is a field WD located near the
cluster.  Accurate, precise proper motion measurements for this WD
will likely be necessary to clarify its cluster membership, and a
higher signal-to-noise spectrum is needed to determine its \teff\ and
\logg.  Clearly, this star warrants further study.

\section{Conclusions\label{sec.concl}}
In this paper, we have presented observations and analysis of the
white dwarf population of the open star cluster M35.  Our conclusions
are the following.

\begin{itemize}

\item Spectroscopy of 14 white dwarf candidates identifies 12 DA, 1
DB, and 1 hot DQ WDs.  Temperatures, surface gravities, masses, and
cooling ages are derived for each of the DA WDs.  

\item All twelve DAs are potentially cluster member WDs, with
  distance moduli consistent with that of the cluster and cooling ages
  less than the cluster age.  The hot DQ and DB are also 
  consistent with being cluster members.  Further data, such as proper
  motion measurements, are necessary to confirm the cluster
  membership of each star.

\item The empirical initial-final mass relation from the M35 white
  dwarf population is consistent with the roughly linear relation
  derived from other open star clusters.

\item The dominant systematic uncertainty in the empirical
initial-final mass relation of M35 is the uncertainty in the star
cluster age.  In the absence of tighter age constraints, we cannot
draw robust conclusions on the intrinsic scatter and
metallicity-dependence of the initial-final mass relation.

\item Based on M35 WDs alone, the lower limit on the maximum mass of
WD progenitor stars (\mcrit) is $\sim 5.1\msun$ (95\%
confidence). Inclusion of WDs from other young open clusters raises this lower
limit to $\sim 6.3\msun - 7.1\msun$, depending on the membership of
certain massive white dwarfs and the core composition of the most
massive WDs.  Combined with upper limits on \mcrit\ from supernova
surveys, $6\msun \lesssim \mcrit \lesssim 9.5\msun$ for
solar-metallicity stars.

\item Based on the cluster WDs alone, we derive a distance modulus to
M35 of $(m-M)_V=10.45\pm 0.08$ and reddening of $\ebv=0.185\pm0.010$,
both in agreement with published values derived from main sequence
fitting.

\item Four of the DAs have masses that are sufficiently massive
  that they may possess oxygen-neon cores. The available data are
  incapable of determining the core compositions, and the core
  composition has only a small impact on the initial mass of these
  stars.

\end{itemize}

\acknowledgements
K.A.W.~ is grateful for the financial support of National Science
Foundation award AST-0602288.  K.A.W.~ and M.B.~ are also grateful for
the support for this project in the form of National Science
Foundation grant AST-0397492.

We wish to thank Jim Liebert and Mike Montgomery for extended
discussions crucial to the results in this paper.  We are grateful for
discussions with Patrick Dufour concerning the hot DQ.  We also thank
Paul Dobbie for helpful discussions and for providing advance access
to his team's newest results. Antonio Kanaan calculated magnetic DA
spectra used to estimate the magnetic field strength of LAWDS 22.  Ted
von Hippel, Don Winget, Elizabeth Jeffery, and Seth Redfield
participated discussions that contributed to this paper in numerous
ways. We wish to thank Matt Wood, Pierre Bergeron, Gilles Fontaine,
L.~G.\ Althaus and their collaborators for making their evolutionary
and photometric models available to us.  This research has made use of
the WEBDA database, operated at the Institute for Astronomy of the
University of Vienna, and the SIMBAD database, operated at CDS,
Strasbourg, France.

The authors wish to recognize and acknowledge the very significant
cultural role and reverence that the summits of Mauna Kea and Kitt
Peak have within the indigenous Hawaiian community and Tohono O'odham
Nation.  We are most fortunate to have the opportunity to conduct
observations from these mountains.

{\it Facilities:} \facility{Mayall (Mosaic-1)}, \facility{Keck:I
  (LRIS-B)}, \facility{Shane (PFCam)}

\bibliographystyle{apj} 

\begin{thebibliography}{96}
\expandafter\ifx\csname natexlab\endcsname\relax\def\natexlab#1{#1}\fi

\bibitem[{{Althaus} {et~al.}(2007){Althaus}, {Garc{\'{\i}}a-Berro}, {Isern},
  {C{\'o}rsico}, \& {Rohrmann}}]{2007A&A...465..249A}
{Althaus}, L.~G., {Garc{\'{\i}}a-Berro}, E., {Isern}, J., {C{\'o}rsico}, A.~H.,
  \& {Rohrmann}, R.~D. 2007, \aap, 465, 249

\bibitem[{{Anders} \& {Grevesse}(1989)}]{1989GeCoA..53..197A}
{Anders}, E., \& {Grevesse}, N. 1989, \gca, 53, 197

\bibitem[{{Anthony-Twarog}(1982)}]{1982ApJ...255..245A}
{Anthony-Twarog}, B.~J. 1982, \apj, 255, 245

\bibitem[{{Barrado y Navascu{\'e}s} {et~al.}(2001{\natexlab{a}}){Barrado y
  Navascu{\'e}s}, {Deliyannis}, \& {Stauffer}}]{2001ApJ...549..452B}
{Barrado y Navascu{\'e}s}, D., {Deliyannis}, C.~P., \& {Stauffer}, J.~R.
  2001{\natexlab{a}}, \apj, 549, 452

\bibitem[{{Barrado y Navascu{\'e}s} {et~al.}(2001{\natexlab{b}}){Barrado y
  Navascu{\'e}s}, {Stauffer}, {Bouvier}, \&
  {Mart{\'{\i}}n}}]{2001ApJ...546.1006B}
{Barrado y Navascu{\'e}s}, D., {Stauffer}, J.~R., {Bouvier}, J., \&
  {Mart{\'{\i}}n}, E.~L. 2001{\natexlab{b}}, \apj, 546, 1006

\bibitem[{{Barstow} {et~al.}(2005){Barstow}, {Bond}, {Holberg}, {Burleigh},
  {Hubeny}, \& {Koester}}]{2005MNRAS.362.1134B}
{Barstow}, M.~A., {Bond}, H.~E., {Holberg}, J.~B., {Burleigh}, M.~R., {Hubeny},
  I., \& {Koester}, D. 2005, \mnras, 362, 1134

\bibitem[{{Barstow} {et~al.}(2003){Barstow}, {Good}, {Burleigh}, {Hubeny},
  {Holberg}, \& {Levan}}]{2003MNRAS.344..562B}
{Barstow}, M.~A., {Good}, S.~A., {Burleigh}, M.~R., {Hubeny}, I., {Holberg},
  J.~B., \& {Levan}, A.~J. 2003, \mnras, 344, 562

\bibitem[{{Barstow} {et~al.}(1998){Barstow}, {Hubeny}, \&
  {Holberg}}]{1998MNRAS.299..520B}
{Barstow}, M.~A., {Hubeny}, I., \& {Holberg}, J.~B. 1998, \mnras, 299, 520

\bibitem[{{Bergeron} {et~al.}(1992){Bergeron}, {Saffer}, \&
  {Liebert}}]{1992ApJ...394..228B}
{Bergeron}, P., {Saffer}, R.~A., \& {Liebert}, J. 1992, \apj, 394, 228

\bibitem[{{Bergeron} {et~al.}(1994){Bergeron}, {Wesemael}, {Beauchamp}, {Wood},
  {Lamontagne}, {Fontaine}, \& {Liebert}}]{1994ApJ...432..305B}
{Bergeron}, P., {Wesemael}, F., {Beauchamp}, A., {Wood}, M.~A., {Lamontagne},
  R., {Fontaine}, G., \& {Liebert}, J. 1994, \apj, 432, 305

\bibitem[{{Bessell}(1995)}]{1995PASP..107..672B}
{Bessell}, M.~S. 1995, \pasp, 107, 672

\bibitem[{{Boesgaard} \& {Friel}(1990)}]{1990ApJ...351..467B}
{Boesgaard}, A.~M., \& {Friel}, E.~D. 1990, \apj, 351, 467

\bibitem[{{Busso} {et~al.}(1999){Busso}, {Gallino}, \&
  {Wasserburg}}]{1999ARA&A..37..239B}
{Busso}, M., {Gallino}, R., \& {Wasserburg}, G.~J. 1999, \araa, 37, 239

\bibitem[{{Casewell} {et~al.}(2008){Casewell}, {Dobbie}, {Napiwotzki},
  {Barstow}, {Burleigh}, \& {Jameson}}]{Casewell2008}
{Casewell}, S.~L., {Dobbie}, P.~D., {Napiwotzki}, R., {Barstow}, M.~A.,
  {Burleigh}, M.~R., \& {Jameson}, R.~F. 2008, \mnras, submitted

\bibitem[{{Catal{\'a}n} {et~al.}(2008){Catal{\'a}n}, {Isern},
  {Garc{\'{\i}}a-Berro}, \& {Ribas}}]{2008MNRAS.387.1693C}
{Catal{\'a}n}, S., {Isern}, J., {Garc{\'{\i}}a-Berro}, E., \& {Ribas}, I. 2008,
  \mnras, 387, 1693

\bibitem[{{Claret}(2007)}]{2007A&A...475.1019C}
{Claret}, A. 2007, \aap, 475, 1019

\bibitem[{{Claver} {et~al.}(2001){Claver}, {Liebert}, {Bergeron}, \&
  {Koester}}]{2001ApJ...563..987C}
{Claver}, C.~F., {Liebert}, J., {Bergeron}, P., \& {Koester}, D. 2001, \apj,
  563, 987

\bibitem[{{Dekel} \& {Silk}(1986)}]{1986ApJ...303...39D}
{Dekel}, A., \& {Silk}, J. 1986, \apj, 303, 39

\bibitem[{{Dobbie} {et~al.}(2006{\natexlab{a}}){Dobbie}, {Napiwotzki},
  {Burleigh}, {Barstow}, {Boyce}, {Casewell}, {Jameson}, {Hubeny}, \&
  {Fontaine}}]{2006MNRAS.369..383D}
{Dobbie}, P.~D., et~al.\  2006{\natexlab{a}}, \mnras, 369, 383

\bibitem[{{Dobbie} {et~al.}(2008){Dobbie}, {Napiwotzki}, {Burleigh},
  {Williams}, {Sharp}, {Barstow}, {Casewell}, \& {Hubeny}}]{Dobbie2008}
{Dobbie}, P.~D., {Napiwotzki}, R., {Burleigh}, M.~R., {Williams}, K., {Sharp},
  R., {Barstow}, M.~A., {Casewell}, S.~L., \& {Hubeny}, I. 2008, \mnras,
  submitted

\bibitem[{{Dobbie} {et~al.}(2006{\natexlab{b}}){Dobbie}, {Napiwotzki},
  {Lodieu}, {Burleigh}, {Barstow}, \& {Jameson}}]{2006MNRAS.373L..45D}
{Dobbie}, P.~D., {Napiwotzki}, R., {Lodieu}, N., {Burleigh}, M.~R., {Barstow},
  M.~A., \& {Jameson}, R.~F. 2006{\natexlab{b}}, \mnras, 373, L45

\bibitem[{{Dobbie} {et~al.}(2004){Dobbie}, {Pinfield}, {Napiwotzki}, {Hambly},
  {Burleigh}, {Barstow}, {Jameson}, \& {Hubeny}}]{2004MNRAS.355L..39D}
{Dobbie}, P.~D., {Pinfield}, D.~J., {Napiwotzki}, R., {Hambly}, N.~C.,
  {Burleigh}, M.~R., {Barstow}, M.~A., {Jameson}, R.~F., \& {Hubeny}, I. 2004,
  \mnras, 355, L39

\bibitem[{{Dufour} {et~al.}(2008){Dufour}, {Fontaine}, {Liebert}, {Schmidt}, \&
  {Behara}}]{Dufour2008}
{Dufour}, P., {Fontaine}, G., {Liebert}, J., {Schmidt}, G.~D., \& {Behara}, N.
  2008, \apj, 683, 978

\bibitem[{{Ferrario} {et~al.}(2005){Ferrario}, {Wickramasinghe}, {Liebert}, \&
  {Williams}}]{2005MNRAS.361.1131F}
{Ferrario}, L., {Wickramasinghe}, D., {Liebert}, J., \& {Williams}, K.~A. 2005,
  \mnras, 361, 1131

\bibitem[{{Fontaine} {et~al.}(2001){Fontaine}, {Brassard}, \&
  {Bergeron}}]{2001PASP..113..409F}
{Fontaine}, G., {Brassard}, P., \& {Bergeron}, P. 2001, \pasp, 113, 409

\bibitem[{{Garcia-Berro} {et~al.}(1997){Garcia-Berro}, {Ritossa}, \&
  {Iben}}]{1997ApJ...485..765G}
{Garcia-Berro}, E., {Ritossa}, C., \& {Iben}, I.~J. 1997, \apj, 485, 765

\bibitem[{{Hartman} {et~al.}(2008){Hartman}, {Gaudi}, {Holman}, {McLeod},
  {Stanek}, {Barranco}, {Pinsonneault}, {Meibom}, \&
  {Kalirai}}]{2008ApJ...675.1233H}
{Hartman}, J.~D., et~al.\  2008, \apj, 675, 1233

\bibitem[{{Hendry} {et~al.}(2006){Hendry}, {Smartt}, {Crockett}, {Maund},
  {Gal-Yam}, {Moon}, {Cenko}, {Fox}, {Kudritzki}, {Benn}, \&
  {{\O}stensen}}]{2006MNRAS.369.1303H}
{Hendry}, M.~A., et~al.\  2006, \mnras, 369, 1303

\bibitem[{{Herbig}(1966)}]{1966ZA.....64..512H}
{Herbig}, G.~H. 1966, Zeitschrift fur Astrophysik, 64, 512

\bibitem[{{Herwig}(2005)}]{2005ARA&A..43..435H}
{Herwig}, F. 2005, \araa, 43, 435

\bibitem[{{Holberg} \& {Bergeron}(2006)}]{2006AJ....132.1221H}
{Holberg}, J.~B., \& {Bergeron}, P. 2006, \aj, 132, 1221

\bibitem[{{Jannuzi} {et~al.}(2003){Jannuzi}, {Claver}, \&
  {Valdes}}]{Jannuzi2003}
{Jannuzi}, B.~T., {Claver}, J., \& {Valdes}, F. 2003, {The NOAO Deep Wide-Field
  Survey MOSAIC Data Reductions} (Tucson: NOAO)

\bibitem[{{Jeffries} {et~al.}(2002){Jeffries}, {Totten}, {Harmer}, \&
  {Deliyannis}}]{2002MNRAS.336.1109J}
{Jeffries}, R.~D., {Totten}, E.~J., {Harmer}, S., \& {Deliyannis}, C.~P. 2002,
  \mnras, 336, 1109

\bibitem[{{Jones} \& {Prosser}(1996)}]{1996AJ....111.1193J}
{Jones}, B.~F., \& {Prosser}, C.~F. 1996, \aj, 111, 1193

\bibitem[{{Kalirai} {et~al.}(2007){Kalirai}, {Bergeron}, {Hansen}, {Kelson},
  {Reitzel}, {Rich}, \& {Richer}}]{2007ApJ...671..748K}
{Kalirai}, J.~S., {Bergeron}, P., {Hansen}, B.~M.~S., {Kelson}, D.~D.,
  {Reitzel}, D.~B., {Rich}, R.~M., \& {Richer}, H.~B. 2007, \apj, 671, 748

\bibitem[{{Kalirai} {et~al.}(2003){Kalirai}, {Fahlman}, {Richer}, \&
  {Ventura}}]{2003AJ....126.1402K}
{Kalirai}, J.~S., {Fahlman}, G.~G., {Richer}, H.~B., \& {Ventura}, P. 2003,
  \aj, 126, 1402

\bibitem[{{Kalirai} {et~al.}(2008){Kalirai}, {Hansen}, {Kelson}, {Reitzel},
  {Rich}, \& {Richer}}]{2007arXiv0706.3894K}
{Kalirai}, J.~S., {Hansen}, B.~M.~S., {Kelson}, D.~D., {Reitzel}, D.~B.,
  {Rich}, R.~M., \& {Richer}, H.~B. 2008, \apj, 676, 594

\bibitem[{{Kalirai} {et~al.}(2005{\natexlab{a}}){Kalirai}, {Richer}, {Hansen},
  {Reitzel}, \& {Rich}}]{2005ApJ...618L.129K}
{Kalirai}, J.~S., {Richer}, H.~B., {Hansen}, B.~M.~S., {Reitzel}, D., \&
  {Rich}, R.~M. 2005{\natexlab{a}}, \apjl, 618, L129

\bibitem[{{Kalirai} {et~al.}(2005{\natexlab{b}}){Kalirai}, {Richer}, {Reitzel},
  {Hansen}, {Rich}, {Fahlman}, {Gibson}, \& {von Hippel}}]{2005ApJ...618L.123K}
{Kalirai}, J.~S., {Richer}, H.~B., {Reitzel}, D., {Hansen}, B.~M.~S., {Rich},
  R.~M., {Fahlman}, G.~G., {Gibson}, B.~K., \& {von Hippel}, T.
  2005{\natexlab{b}}, \apjl, 618, L123

\bibitem[{{Kalirai} \& {Tosi}(2004)}]{2004MNRAS.351..649K}
{Kalirai}, J.~S., \& {Tosi}, M. 2004, \mnras, 351, 649

\bibitem[{{Koester} {et~al.}(2001){Koester}, {Napiwotzki}, {Christlieb},
  {Drechsel}, {Hagen}, {Heber}, {Homeier}, {Karl}, {Leibundgut}, {Moehler},
  {Nelemans}, {Pauli}, {Reimers}, {Renzini}, \&
  {Yungelson}}]{2001A&A...378..556K}
{Koester}, D., et~al.\  2001, \aap, 378, 556

\bibitem[{{Koester} \& {Reimers}(1996)}]{1996A&A...313..810K}
{Koester}, D., \& {Reimers}, D. 1996, \aap, 313, 810

\bibitem[{{Lajoie} \& {Bergeron}(2007)}]{2007ApJ...667.1126L}
{Lajoie}, C.-P., \& {Bergeron}, P. 2007, \apj, 667, 1126

\bibitem[{{Landolt}(1992)}]{1992AJ....104..340L}
{Landolt}, A.~U. 1992, \aj, 104, 340

\bibitem[{{Leonard} \& {Merritt}(1989)}]{1989ApJ...339..195L}
{Leonard}, P.~J.~T., \& {Merritt}, D. 1989, \apj, 339, 195

\bibitem[{{Liebert} {et~al.}(2005{\natexlab{a}}){Liebert}, {Bergeron}, \&
  {Holberg}}]{2005ApJS..156...47L}
{Liebert}, J., {Bergeron}, P., \& {Holberg}, J.~B. 2005{\natexlab{a}}, \apjs,
  156, 47

\bibitem[{{Liebert} {et~al.}(2005{\natexlab{b}}){Liebert}, {Young}, {Arnett},
  {Holberg}, \& {Williams}}]{2005ApJ...630L..69L}
{Liebert}, J., {Young}, P.~A., {Arnett}, D., {Holberg}, J.~B., \& {Williams},
  K.~A. 2005{\natexlab{b}}, \apjl, 630, L69

\bibitem[{{Marigo}(2001)}]{2001A&A...370..194M}
{Marigo}, P. 2001, \aap, 370, 194

\bibitem[{{Marigo} \& {Girardi}(2007)}]{2007A&A...469..239M}
{Marigo}, P., \& {Girardi}, L. 2007, \aap, 469, 239

\bibitem[{{Marigo} {et~al.}(2008){Marigo}, {Girardi}, {Bressan}, {Groenewegen},
  {Silva}, \& {Granato}}]{2008A&A...482..883M}
{Marigo}, P., {Girardi}, L., {Bressan}, A., {Groenewegen}, M.~A.~T., {Silva},
  L., \& {Granato}, G.~L. 2008, \aap, 482, 883

\bibitem[{{Marshall} {et~al.}(2005){Marshall}, {Burke}, {DePoy}, {Gould}, \&
  {Kollmeier}}]{2005AJ....130.1916M}
{Marshall}, J.~L., {Burke}, C.~J., {DePoy}, D.~L., {Gould}, A., \& {Kollmeier},
  J.~A. 2005, \aj, 130, 1916

\bibitem[{{Martin}(2005)}]{2005ApJ...621..227M}
{Martin}, C.~L. 2005, \apj, 621, 227

\bibitem[{{Massey} \& {Slesnick}(1999)}]{Massey1999}
{Massey}, P., \& {Slesnick}, C.~L. 1999, NOAO Newsletter, 59

\bibitem[{{Mazzei} \& {Pigatto}(1989)}]{1989A&A...213L...1M}
{Mazzei}, P., \& {Pigatto}, L. 1989, \aap, 213, L1

\bibitem[{{McCarthy} {et~al.}(1998)}]{1998SPIE.3355...81M}
{McCarthy}, J.~K., {et~al.} 1998, in Proc. SPIE Vol. 3355, p. 81-92, Optical
  Astronomical Instrumentation, Sandro D'Odorico; Ed., ed. S.~{D'Odorico},
  81--92

\bibitem[{{McKee} \& {Ostriker}(1977)}]{1977ApJ...218..148M}
{McKee}, C.~F., \& {Ostriker}, J.~P. 1977, \apj, 218, 148

\bibitem[{{McNamara} \& {Sekiguchi}(1986)}]{1986ApJ...310..613M}
{McNamara}, B.~J., \& {Sekiguchi}, K. 1986, \apj, 310, 613

\bibitem[{{Mermilliod}(1981)}]{1981A&A....97..235M}
{Mermilliod}, J.~C. 1981, \aap, 97, 235

\bibitem[{{Meynet} {et~al.}(1993){Meynet}, {Mermilliod}, \&
  {Maeder}}]{1993A&AS...98..477M}
{Meynet}, G., {Mermilliod}, J.-C., \& {Maeder}, A. 1993, \aaps, 98, 477

\bibitem[{{Mochejska} {et~al.}(2004){Mochejska}, {Stanek}, {Sasselov},
  {Szentgyorgyi}, {Westover}, \& {Winn}}]{2004AJ....128..312M}
{Mochejska}, B.~J., {Stanek}, K.~Z., {Sasselov}, D.~D., {Szentgyorgyi}, A.~H.,
  {Westover}, M., \& {Winn}, J.~N. 2004, \aj, 128, 312

\bibitem[{{Napiwotzki}(1997)}]{1997A&A...322..256N}
{Napiwotzki}, R. 1997, \aap, 322, 256

\bibitem[{{Napiwotzki} {et~al.}(1999){Napiwotzki}, {Green}, \&
  {Saffer}}]{1999ApJ...517..399N}
{Napiwotzki}, R., {Green}, P.~J., \& {Saffer}, R.~A. 1999, \apj, 517, 399

\bibitem[{{Oke} {et~al.}(1995)}]{1995PASP..107..375O}
{Oke}, J.~B., {et~al.} 1995, \pasp, 107, 375

\bibitem[{{Origlia} {et~al.}(2006){Origlia}, {Valenti}, {Rich}, \&
  {Ferraro}}]{2006ApJ...646..499O}
{Origlia}, L., {Valenti}, E., {Rich}, R.~M., \& {Ferraro}, F.~R. 2006, \apj,
  646, 499

\bibitem[{{Perryman} {et~al.}(1998)}]{1998A&A...331...81P}
{Perryman}, M.~A.~C., {et~al.} 1998, \aap, 331, 81

\bibitem[{{Pinsonneault} {et~al.}(1998){Pinsonneault}, {Stauffer}, {Soderblom},
  {King}, \& {Hanson}}]{1998ApJ...504..170P}
{Pinsonneault}, M.~H., {Stauffer}, J., {Soderblom}, D.~R., {King}, J.~R., \&
  {Hanson}, R.~B. 1998, \apj, 504, 170

\bibitem[{{Poelarends} {et~al.}(2008){Poelarends}, {Herwig}, {Langer}, \&
  {Heger}}]{2008ApJ...675..614P}
{Poelarends}, A.~J.~T., {Herwig}, F., {Langer}, N., \& {Heger}, A. 2008, \apj,
  675, 614

\bibitem[{{Poelarends} {et~al.}(2006){Poelarends}, {Izzard}, {Herwig},
  {Langer}, \& {Heger}}]{2006MmSAI..77..846P}
{Poelarends}, A.~J.~T., {Izzard}, R.~G., {Herwig}, F., {Langer}, N., \&
  {Heger}, A. 2006, Memorie della Societa Astronomica Italiana, 77, 846

\bibitem[{{Press} {et~al.}(1992){Press}, {Teukolsky}, {Vetterling}, \&
  {Flannery}}]{Recipes}
{Press}, W.~H., {Teukolsky}, S.~A., {Vetterling}, W.~T., \& {Flannery}, B.~P.
  1992, {Numerical recipes in FORTRAN. The art of scientific computing}
  (Cambridge: University Press, |c1992, 2nd ed.)

\bibitem[{{Reimers} \& {Koester}(1988)}]{1988A&A...202...77R}
{Reimers}, D., \& {Koester}, D. 1988, \aap, 202, 77

\bibitem[{{Rieke} \& {Lebofsky}(1985)}]{1985ApJ...288..618R}
{Rieke}, G.~H., \& {Lebofsky}, M.~J. 1985, \apj, 288, 618

\bibitem[{{Romanishin} \& {Angel}(1980)}]{1980ApJ...235..992R}
{Romanishin}, W., \& {Angel}, J.~R.~P. 1980, \apj, 235, 992

\bibitem[{{Rubin} {et~al.}(2008){Rubin}, {Williams}, {Bolte}, \&
  {Koester}}]{Rubin2007}
{Rubin}, K.~H.~R., {Williams}, K.~A., {Bolte}, M., \& {Koester}, D. 2008, \aj,
  135, 2163

\bibitem[{{Salaris} {et~al.}(2008){Salaris}, {Serenelli}, {Weiss}, \& {Miller
  Bertolami}}]{2008arXiv0807.3567S}
{Salaris}, M., {Serenelli}, A., {Weiss}, A., \& {Miller Bertolami}, M. 2008, \apj, submitted (arXiv:0807.3567)

\bibitem[{{Salpeter}(1955)}]{1955ApJ...121..161S}
{Salpeter}, E.~E. 1955, \apj, 121, 161

\bibitem[{{Sandberg Lacy} {et~al.}(2008){Sandberg Lacy}, {Torres}, \&
  {Claret}}]{2008AJ....135.1757S}
{Sandberg Lacy}, C.~H., {Torres}, G., \& {Claret}, A. 2008, \aj, 135, 1757

\bibitem[{{Sarajedini} {et~al.}(2004){Sarajedini}, {Brandt}, {Grocholski}, \&
  {Tiede}}]{2004AJ....127..991S}
{Sarajedini}, A., {Brandt}, K., {Grocholski}, A.~J., \& {Tiede}, G.~P. 2004,
  \aj, 127, 991

\bibitem[{{Schuler} {et~al.}(2003){Schuler}, {King}, {Fischer}, {Soderblom}, \&
  {Jones}}]{2003AJ....125.2085S}
{Schuler}, S.~C., {King}, J.~R., {Fischer}, D.~A., {Soderblom}, D.~R., \&
  {Jones}, B.~F. 2003, \aj, 125, 2085

\bibitem[{{Siess}(2007)}]{2007A&A...476..893S}
{Siess}, L. 2007, \aap, 476, 893

\bibitem[{{Slesnick} {et~al.}(2002){Slesnick}, {Hillenbrand}, \&
  {Massey}}]{2002ApJ...576..880S}
{Slesnick}, C.~L., {Hillenbrand}, L.~A., \& {Massey}, P. 2002, \apj, 576, 880

\bibitem[{{Smartt} {et~al.}(2008){Smartt}, {Eldridge}, {Crockett}, \&
  {Maund}}]{Smartt2008}
{Smartt}, S.~J., {Eldridge}, J.~J., {Crockett}, R.~M., \& {Maund}, J.~R. 2008,
  \mnras, submitted (arXiv:0809.0403)

\bibitem[{{Stauffer} {et~al.}(1998){Stauffer}, {Schultz}, \&
  {Kirkpatrick}}]{1998ApJ...499L.199S}
{Stauffer}, J.~R., {Schultz}, G., \& {Kirkpatrick}, J.~D. 1998, \apjl, 499,
  L199

\bibitem[{{Stetson}(1987)}]{1987PASP...99..191S}
{Stetson}, P.~B. 1987, \pasp, 99, 191

\bibitem[{{Sung} \& {Bessell}(1999)}]{1999MNRAS.306..361S}
{Sung}, H., \& {Bessell}, M.~S. 1999, \mnras, 306, 361

\bibitem[{{Sung} {et~al.}(2002){Sung}, {Bessell}, {Lee}, \&
  {Lee}}]{2002AJ....123..290S}
{Sung}, H., {Bessell}, M.~S., {Lee}, B.-W., \& {Lee}, S.-G. 2002, \aj, 123, 290

\bibitem[{{Tout} {et~al.}(2008){Tout}, {Wickramasinghe}, {Liebert}, {Ferrario},
  \& {Pringle}}]{2008MNRAS.387..897T}
{Tout}, C.~A., {Wickramasinghe}, D.~T., {Liebert}, J., {Ferrario}, L., \&
  {Pringle}, J.~E. 2008, \mnras, 387, 897

\bibitem[{{Twarog} {et~al.}(1997){Twarog}, {Ashman}, \&
  {Anthony-Twarog}}]{1997AJ....114.2556T}
{Twarog}, B.~A., {Ashman}, K.~M., \& {Anthony-Twarog}, B.~J. 1997, \aj, 114,
  2556

\bibitem[{{van Dokkum}(2001)}]{2001PASP..113.1420V}
{van Dokkum}, P.~G. 2001, \pasp, 113, 1420

\bibitem[{{von Hippel}(2005)}]{2005ApJ...622..565V}
{von Hippel}, T. 2005, \apj, 622, 565

\bibitem[{{von Hippel} {et~al.}(2002){von Hippel}, {Steinhauer}, {Sarajedini},
  \& {Deliyannis}}]{2002AJ....124.1555V}
{von Hippel}, T., {Steinhauer}, A., {Sarajedini}, A., \& {Deliyannis}, C.~P.
  2002, \aj, 124, 1555

\bibitem[{{Weidemann}(2000)}]{2000A&A...363..647W}
{Weidemann}, V. 2000, \aap, 363, 647

\bibitem[{{Wickramasinghe} \& {Ferrario}(2000)}]{2000PASP..112..873W}
{Wickramasinghe}, D.~T., \& {Ferrario}, L. 2000, \pasp, 112, 873

\bibitem[{{Williams} \& {Bolte}(2007)}]{2007AJ....133.1490W}
{Williams}, K.~A., \& {Bolte}, M. 2007, \aj, 133, 1490

\bibitem[{{Williams} {et~al.}(2004){Williams}, {Bolte}, \&
  {Koester}}]{2004ApJ...615L..49W}
{Williams}, K.~A., {Bolte}, M., \& {Koester}, D. 2004, \apjl, 615, L49

\bibitem[{{Williams} {et~al.}(2006){Williams}, {Liebert}, {Bolte}, \&
  {Hanson}}]{2006ApJ...643L.127W}
{Williams}, K.~A., {Liebert}, J., {Bolte}, M., \& {Hanson}, R.~B. 2006, \apjl,
  643, L127

\bibitem[{{Wood}(1995)}]{1995LNP...443...41W}
{Wood}, M.~A. 1995, LNP Vol.~443: White Dwarfs, 443, 41

\end{thebibliography}

\clearpage

\clearpage

\clearpage

\clearpage

\clearpage

\clearpage
\end{document}